\documentclass[a4paper,12pt]{article}
\usepackage{jheppub}
\usepackage{url}

\usepackage{graphicx}
\usepackage{subcaption}
\usepackage{rotating}
\usepackage[T1]{fontenc}
\usepackage{multirow}
\usepackage[]{algorithm2e}
\usepackage{indentfirst}
\usepackage{placeins}
\usepackage{slashed}
\usepackage{gensymb}
\usepackage{amsmath}
\usepackage{amssymb}
\usepackage{feynmp-auto}
\usepackage{multirow}
\usepackage{changes}
\definechangesauthor[color=red]{yu}

\def\Acknowledgements{\bigskip  \bigskip \begin{center} \begin{large}
             \bf ACKNOWLEDGEMENTS \end{large}\end{center}}
\def\APPENDIX{\bigskip  \bigskip {\begin{center}\Large\bf APPENDIX \end{center}}}

\author[a,b]{Sascha Caron}
\emailAdd{scaron@nikhef.nl}
\author[c]{\\Roberto Ruiz de Austri}
\emailAdd{rruiz@ific.uv.es}
\author[a,b]{\\Zhongyi Zhang}
\emailAdd{zzhang@nikhef.nl}

\affiliation[a]{High Energy Physics, Radboud University Nijmegen, Heyendaalseweg 135, 6525 AJ Nijmegen, the Netherlands}
\affiliation[b]{Nikhef, Science Park 105, 1098 XG Amsterdam, the Netherlands}
\affiliation[c]{Instituto de F\'isica Corpuscular, IFIC-UV/CSIC, Valencia, Spain}

\title{Mixture--of--theories Training:\\ Can We Find New Physics and Anomalies Better by Mixing Physical Theories?}

\abstract{
Model--independent search strategies have been increasingly proposed in recent years because on the one hand there has been no clear signal for new physics and on the other hand there is a lack of a highly probable and parameter--free extension of the standard model. For these reasons, there is no simple search target so far.
In this work, we try to take a new direction and ask the question: bearing in mind that we have a large number of new physics theories that go beyond the Standard Model and may contain a grain of truth,
 can we improve our search strategy for unknown signals by using them ``in combination''?
In particular, we show that a signal hypothesis based on a large, intermingled set of
many different theoretical signal models can be a superior approach to find an unknown BSM signal.
Applied to a recent data challenge, we show that ``mixture--of--theories training'' outperforms  strategies that optimize signal regions with a single BSM model as well as most unsupervised strategies.
Applications of this work include anomaly detection and the definition of signal regions in the search for signals of new physics.
}

\begin{document}

\maketitle

\section{Introduction}
\label{sec:intro}

There are several phenomena that are not described by our model of the subatomic world, the Standard Model (SM) of particles.
 These phenomena include dark matter, baryon asymmetry, neutrino masses and a description of gravity in regions where quantum effects cannot be ignored. It is therefore inevitable that the Standard Model must be extended and these extensions are likely to yield new experimental phenomena. We call them signals from physics
beyond the Standard Model of particle physics (BSM).

Without a conclusive set of deviations from the Standard Model (SM) or a new ground--breaking theoretical insight, however, it is unclear which specific BSM signals can be expected. 
With none of the theories proposed over the past few decades proven to be correct, a more data--driven search for signals is becoming increasingly important.

As a result, many BSM theories have been proposed in recent decades, and in addition, these models have many free parameters. 
The characteristic (high--dimensional) distributions (e.g. particle energies, particle types, decay angles) of signal events, measured at a collider or astroparticle experiment, change depending on the different theoretical models and parameters. This leads to enormous possibilities of how a distribution of signal events, might look like.

An important question is therefore:
Without knowing the true BSM theory and the true signal distribution for new physics, how can we further improve our search strategy for signals of new physics?

One way is to follow the traditional approach of performing many searches, each search optimizing the data selection for a given signal.
In particle physics, multivariate methods for optimizing data selection have been used for many years by dividing events with features $x$ into signal--like events and background--like events. Approaches to creating classifiers include kernel methods, decision trees, or neural networks.

Attempts to train a classifier on high--dimensional features to select a data sample where the signal is enhanced have existed since 30 years (or more), see e.g. \cite{Peterson:1993nk,Stimpfl-Abele:1993xgr,Ametller:1996ri, Denby:1987rk}. Typically statistical tests (goodness of fit of the SM hypothesis, hypothesis test for the BSM hypothesis) are done with a selection on the classifier output, i.e. $y> y_{cut}$ or by fitting a signal model on the 1--dimensional classifier output distribution, e.g. \cite{CDF:2009itk,ALEPH:1998tek,ATLAS:2018gfm}.
This selection of data is referred to as the ``signal region'', since the selection enhances the ratio of signal events if the selected BSM model is true. Although such approaches work when the signal is known (prominent examples are the top quark or Higgs discovery), a problem arises when the signal is unknown (or only vaguely known). The signal regions optimized for a certain BSM model (and parameter set) might -- in real data -- not select the right signals.

Proposals to overcome this problem were to move away from concrete BSM models and define more generic models (e.g. simplified models or effective field theories) or to look for new physics without a concrete signal model (model independent).
However also a simplified model or a particular effective theory model is just another model with physical parameters and it could still not yield the right signal phenomenology. Another increasingly popular method for finding new physics is via so–called unsupervised Machine Learning (ML) methods.  Here the idea is to completely neglect the signal hypothesis and use only data or
the standard model hypothesis, i.e. the known physics, to search for new physics.

In this work we  try to take a different direction and ask the question:
Given that we have a large number of new physical models and that there may be some truth in them, how can we further improve our search strategy for signals from new physics?
In particular, we test whether a signal hypothesis based on a large, intermingled set of many different theoretical signal models is a good approach to find an unknown BSM signal.

We compare this strategy with supervised approaches and with a large number of unsupervised approaches based on the datasets and results found in the DarkMachines data challenge ~\cite{Aarrestad:2021oeb}. 

It is important to emphasize that this comparison assumes that the ``data signal'' is unknown, just like the real data. The performance of the compared methods is calculated with about 20 data sets with different BSM signals and is additionally calculated with a ``secret'' (i.e. completely blinded) dataset that contains an unknown number and type of signal events. In addition, we use 4 different key data selections (channels), including 2 hadronic final states and 2 leptonic final states. All channels have a requirement on missing transverse momentum and are defined in ~\cite{Aarrestad:2021oeb}.

The study is grouped as follows. First, we  discuss the 3 different strategies for training the classifier in  section~\ref{sec:strategies}. Second, we introduce the training dataset we used for our supervised strategy and test dataset for the comparison in  section~\ref{sec:data}. Third, we benchmark ML--models from the supervised strategy trained on different dataset to analyze the performance tested on the {\tt hackathon data} in  section~\ref{sec:supervised}. Fourth, the supervised methods are compared with unsupervised methods tested on both {\tt hackathon data} and {\tt secret data} in the section~\ref{sec:unsupervised}. Finally, we summarize and conclude our research in section~\ref{sec:results}.

\section{Three Strategies to Train Networks for Searching New Physics:\\ (Single) BSM Models, Model--independent,\\ and Mixture of Theories}
\label{sec:strategies}

\subsection{Signal Regions Defined by (Single) BSM Models}

One aim of theoretical and phenomenological research is to develop testable BSM models and to determine the expected parameter ranges of these models. 
The phenomenology of these BSM models is determined through simulations and then provides hypothetical signals for new physics such as collision events. A BSM model is tested for true or false with an experimental data sample enriched with the BSM signal events expected from the simulation.

The specific details of this statistical test can be found elsewhere in the literature \cite{Read:2002hq,Cowan:2010js} and are not relevant for this work. In this work we use the figures--of--merit defined Ref.~\cite{Aarrestad:2021oeb}. Here the signal improvement (SI) is defined as the ratio of the selection efficiency of the signal events divided by the square root of the selection efficiency of the background events. Data selection that improves SI is expected to yield better sensitivity to the hypothetical BSM signal. We assume that the SM prediction is known from the simulation or from the auxiliary measurement and that interfering nuisance parameters can be included in the statistical test. To determine these SM background predictions, control measurements with so--called control regions are used in particular, see e.g. \cite{ATLAS:2011xeq,ATLAS:2011hnu}.

The data selection that optimizes signal sensitivity (also referred to as the ``signal region'' at LHC) is often defined using a specific observable 1--dimensional event feature such as a reconstructed particle mass or through an ML--based classification output.
The classifier is typically trained to separate a signal hypothesis for a signal $S$ (often marked in the training as $1$) that can be described with a distribution $p_{S}(x)$ from the SM events given by $p_{SM}(x)$ (marked as $0$).  These features $x$ can be the 4--vectors of many particles, so the problem is intrinsically  high--dimensional. 
 The training is done with the help of simulated data: many signal and SM events are simulated and would follow the  true distributions $p_S(x)$ and $p_{SM}(x)$ for infinite statistics. Training means that the parameters of the classifier (e.g. the cuts of the decision trees or the weights of a neural network) are optimized with the training data by using an objective (or loss) function, e.g. the binary cross entropy or Gini impurity is minimized.

If the tested BSM signal is correct or we at least choose a BSM model (and parameter set) very similar to the one realized in nature, the classifier should enhance the signal in real data at high classifier output scores. In contrast, if the real world signal is quite different from the assumed BSM signal, the classifier will not improve (or even decrease) the signal sensitivity when applied to the real data.
In this work, we test how well this supervised single--model optimization strategy performs in a quasi--realistic environment, i.e. to find unknown signals that are not the ones used to train the classifier. We call this
``Single Signal Approach (SSA)''. The detailed training process used for this strategy is described in subsection~\ref{SSA}.
There are some proposed extensions to this simple optimization procedure that make classifiers more agnostic, e.g. in relation to the width or mass of a given signal in a given BSM model, see
 Ref.~\cite{Baldi:2016fzo,Drees:2021oew}.
In this study, however, we only compare with the widely used simple SSA methods.

\subsection{Signal Regions Defined by Model--independent, Data--derived or Unsupervised Strategies}
Signal regions have also been defined without the assumption of a signal model.
One method is to test an enormous amount of signal regions using an automated procedure. The disadvantage of these methods is the automated determination of the background and the problem of multiple testing, which can reduce the significance.
Such \textit{general} searches without an explicit BSM signal assumption have been performed by the D\O\ Collaboration~\cite{Abbott:2000fb,Abbott:2000gx,Abbott:2001ke, Abazov:2011ma} at the Tevatron,
by the H1 Collaboration~\cite{Aktas:2004pz,Aaron:2008aa} at HERA, by the CDF Collaboration~\cite{Aaltonen:2007dg, Aaltonen:2008vt} at the Tevatron and scanning thousands of analysis channels at the LHC by ATLAS and CMS comparing data to SM simulations~\cite{Aaboud:2018ufy,Sirunyan:2020jwk}. Ref.~\cite{Aaboud:2018ufy} explicitly suggests defining ``data--derived signal regions'' via the excesses found and testing these excesses with new data and standard approaches, e.g. by determining a background model with auxiliary measurements and control regions.

Other approaches are based on unsupervised (or weakly supervised) classifiers. Machine Learning algorithms are trained (unsupervised, i.e. without a signal model) to estimate the (high--dimensional) distribution $p_{SM}(x)$ of SM events with event topology or event features $x$ or to learn to classify events as stemming from the SM prediction. 
Unsupervised refers to the fact that no signal data is used to supervise the ML--model. Many approaches are based on AutoEncoders that map the data defined by a set of features $x$ to a code $z$ and then map the code again to data features $x'$.
Compression--decompression is learned with training data and the AutoEncoder is assumed to be less good at reconstructing anomalies (i.e. data with characteristics different from those used to train the AutoEncoder).
In the most extreme unsupervised cases, only the distribution $p_{SM}(x)$ as expected of the Standard Model is needed \cite{Caron:2021wmq}. The Standard Model prediction can come from simulated events, can be directly taken from data assuming that a signal is small or can be estimated from auxiliary measurements (outside of a signal data selection) using assumptions about how the distribution of SM events behaves in the signal selection.
Techniques that propose unsupervised Deep Learning to search for new physics are e.g. \cite{Wozniak:2020cry,Knapp:2020dde,Dery:2017fap,Cohen:2017exh,Metodiev:2017vrx,DAgnolo:2018cun,Heimel:2018mkt,Farina:2018fyg,Komiske:2018oaa,Hajer:2018kqm,Cerri:2018anq,Collins:2018epr,Otten:2019hhl,Blance:2019ibf,Collins:2019jip,Sirunyan:2019jud,vanBeekveld:2020txa,Aad:2020cws,Nachman:2020lpy,Kasieczka:2021xcg,Caron:2021wmq}. Comparisons of unsupervised approaches were performed in Ref.~\cite{Aarrestad:2021oeb} and we use these datasets and the results found to compare our supervised strategies with unsupervised approaches.
Another comparison of model--independent approaches using 3 different black boxes based on density comparisons (and bump hunting) between data and expectation is the LHC Olympics ~\cite{Kasieczka:2021xcg}.

\subsection{Signal Regions Defined by ``Mixture of Theories'' or a Hypertheory}
In this work we try to take a different direction and ask the question:
Given that we have a large number of new physical models and that there may be some truth in them, how can we further improve our search strategy for signals from new physics?
Here we test whether a large intermingled set of many different theoretical signal models might have better predictive power.
We can assume that perhaps not all of the proposed model/parameter sets are completely wrong, but that they contain some knowledge about the new phenomenon that we expect to find in the data. This can be considered as building a prior based on the BSM theory community. It is not clear a priori whether this is really a good approach: it may be that the wide range of choices greatly reduces the sensitivity to a particular signal.

\subsubsection{How  Mixture of Theories (MoT) works}
Concretely, we propose to train a  classifier on a {\it mixture of the knowledge of the theoretical community}, i.e. a signal model given by a mixture of many signals with a density that is the mixture distribution $\sum_i^N w_i p_{S,i}(x)$. Here we mix $N$ signal models that can come from completely different theories (e.g. mixing supersymmetry and simplified models), particle sets (mixing Z' with gluinos) and parameter sets. 
Specifically, we use a supervised classifier trained to separate this ``wild mix of signals''  from a SM distribution $p_{SM}$. Since classifiers learn density ratios, the output of the MoT classifier can be interpreted as the ratio of the density of the mixture of BSM models and the SM density.

Since we do not want to favor any of the models, we use $w_i=1$ in this study for all models $i$. Our approach might be extended with more sophisticated approaches to pick training data, e.g. based on active learning~\cite{Caron:2019xkx}, with a mixture of signal events produced by a generative ML model~\cite{Otten:2019hhl} or by using model parameters in the proximity of a learned high--dimensional exclusion boundary \cite{Caron:2016hib}.

To avoid a special design for the training data set, we randomly choose a large number of signals from different BSM models  with multiple parameter settings. When selecting those points we used three criteria:
\begin{itemize}
\item
BSM model prior: We took ``typical models'' used in BSM phenomenology (e.g. SUSY, simplified models etc.)
\item
Variety and Efficiency: We selected the models to have a good coverage of possible BSM signals (e.g. yielding not only signals producing jets and no leptons) and a reasonable efficiency for one of our four signal channels
\item 
BSM parameter prior: We took (random) parameter sets around the expected LHC sensitivity limit (and avoided excluded scenarios).

\end{itemize}
We have carefully tried to avoid selecting models or points based on the test dataset, and we are also unaware of the {\tt DarkMachines Secret Data} BSM models that we also use to validate this approach.

In total we selected 31 different signal models (i.e. producing different BSM particles and different decays), and each with different parameter configurations (e.g. mass combinations). In total, we take $N$ signals from a total set of 116 signals to create the training data sets.

The dataset for training with the mixture of theories (MoT) strategy is described in subsection~\ref{MoTdata}, while the details for training with MoT strategy are discussed in subsection~\ref{MoT}. The comparison between MoT and SSA are presented in subsection~\ref{MoTvsSSA}.

\section{Dataset for Training and Comparison}
\label{sec:data}

\subsection{Training Data: Various Signals for New Physics}
\label{MoTdata}

To train the ML models used to classify SM and BSM events, we used the SM events from the {\tt hackathon data} (see Section 2 in Ref. \cite{Aarrestad:2021oeb}).

The signals we use in this paper come from two types of models. The first are simplified models that add limited sectors to the SM and could be part of the whole model, or their approximation. In contrast to the effective models, the simplified models contain only renormalizable terms without integrating out  physical particles. Second, we use a complete model with enormous phenomenological diversity such as the Minimal Supersymmetric Standard Model (MSSM) \cite{MARTIN_1998}. In this context, we consider several realisations, including its extension with R--parity broken terms (RPV) in the superpotential, motivated by the rich phenomenology that this model would provide in a collider experiment. As for the New Physics signals at the LHC, the different sectors of the MSSM can mimic the signals expected from a diversity of non--SUSY models. This is because in the final state you can have jets, leptons, photons or combinations with large multiplicities, depending on the region of the model's parameter space, plus missing energy.

As mentioned in section 2, we have generated 116 BSM signals containing a variety of phenomena. In this way, we ensure that our method is not specifically designed for a particular task, but is a general method for detecting new physics. Next, we used the selection rules from Ref. \cite{Aarrestad:2021oeb} to create four different channels. Namely, channel 1 (typical SUSY channel with 4 jets and missing transverse momentum), channel 2a (three leptons and missing transverse momentum), channel 2b (two leptons and missing transverse momentum) and channel 3 (trigger level analysis with cut on transverse momentum and missing transverse momentum).

In the following subsections, we present details of the signals that we used for the training process, i.e. only the signals that survive from the event selection of the given channels. The details of the file names of the signals can be found in Appendix~\ref{app:data}.

The signals are coded in either {\tt SARAH} \cite{Staub:2013tta} or {\tt FeynRules} \cite{Alloul:2013bka}, and stored in {\tt UFO} format \cite{Degrande:2011ua} for the simulation by {\tt MadGraph 5} \cite{Alwall:2011uj} to generate parton level events. The parton showers in addition to the hadronization were performed with 
{\tt PYTHIA 8} \cite{Sjostrand:2014zea} and the detector simulation was done with {\tt DELPHES 3} \cite{deFavereau:2013fsa}. We stored the resulting events in {\tt ROOT} files and finally we applied some requirements, described in Sec. 2 of Ref. ~\cite{Aarrestad:2021oeb}, to the resulting objects event by event. The events that met the requiriments were stored in {\tt CSV} text files and made available \cite{darkmachines_community_2022_6772776}.

\subsubsection{Signals from Simplified Models}

For simplified models with new particles, the selection of coupling constants only affects the decay width of on--shell particles. Therefore, we selected only the mass parameters and set all new coupling parameters to 1. Even if the signals we selected are already excluded for this coupling, we could re--scale the couplings to small values to avoid the exclusion. 
Such BSM models are therefore still interesting test signals. We have generated signal samples for the following simplified models:

\begin{enumerate}

\item {\tt btzpLoop} is the physical model contain a leptophobic $U(1)$ mediator $Z^\prime$ that only couples to $b$ quarks and $t$ quarks. The motivation of the model is that the universal couplings between mediator and quarks has the effect that the couplings between mediator and $u$ quark or $d$ quark govern the phenomena for direct detection and LHC detection, since $u$ and $d$ are valence quarks. After avoiding the coupling with light quarks, the box diagram for $gg\rightarrow gZ^\prime$ through a quark loop has comparable contributions to tree diagrams. We generate events for both tree and loop diagrams, but only loop diagram survive from the selection. More details can be found in  \cite{Drees:2018gvr, Drees:2019iex}. This is the only model where we consider loop effects.

\item {\tt LReff}\footnote{http://feynrules.irmp.ucl.ac.be/wiki/EffLRSM} is the physical model describing effective left--right symmetric gauge boson $W^{\prime\pm}$ and $Z^\prime$ coupled to leptons and quarks. More details can be find in the reference \cite{Mattelaer:2016ynf}.

\item {\tt type2seesaw}\footnote{http://feynrules.irmp.ucl.ac.be/wiki/TypeIISeesaw} is the physical model describing a type--2 seesaw mechanism with one particle from the adjoint representation of the weak group carrying hypercharge equal to $1$.   Spontaneous symmetry breaking yields one extra CP--even Higgs, one charged scalar, and one doubly charged scalar particle. More details can be find in the reference \cite{Fuks:2019clu}. 

\item {\tt Wprime}\footnote{https://feynrules.irmp.ucl.ac.be/wiki/Wprime} is the physical model describing a $W^\prime$ coupled to leptons and quarks. More details can be find in the reference \cite{Duffty:2012rf}. 

\item {\tt mDM}\footnote{https://feynrules.irmp.ucl.ac.be/wiki/MDMmodel} is the physical model describing the minimal dilaton model contains a linearized dilation field S. More details can be find in the reference \cite{Abe:2012eu}. 

\end{enumerate}

\subsubsection{Signals from SUSY Models}

As mentioned earlier, we have selected signals based on the MSSM with/without RPV. We have also assumed that the lightest neutralino is the lightest SUSY particle, except for model points with RPV and in points of the Gauge Mediation SUSY Breaking (GMSB) model \cite{Giudice_1999}. In the following description of the signals we refer to squarks as the first and second generation ones, which we assume to have a common mass. For the third generation we consider only the lightest stops and sbottoms. 

We have generated the following signals:

\begin{enumerate}
\item {\tt HH} corresponds to the production of a pair of charged Higgses ($H^\pm$). When the $H^\pm$ is heavier than the top--quark, it decays to a top--quark and a b--quark. Otherwise it decays to a lepton and a neutrino.

\item {\tt hh} corresponds to the production of a pair of heavy CP--even Higgses ($H$) decaying to a pair of W--bosons. 
    
\item {\tt AA} corresponds to the production of a pair of heavy CP--odd Higgses ($A$) decaying mostly to a pair of b--quarks.     
    
\item {\tt glgl} corresponds to gluino ($\tilde{g}$) pair production with a long decay involving electro--weakinos which decay to gauge bosons 
and the lightest neutralino ($\tilde{\chi}^0_1$). The gauge bosons decay inclusively. 
    
\item {\tt sqsq} corresponds to squark ($\tilde{q}$) pair production decaying with a long decay chain ending in light--quarks plus a $\tilde{\chi}^0_1$. 

\item {\tt glsq} corresponds to $\tilde{g} \tilde{q}$ production both decaying to the lightest chargino $\tilde{\chi}^\pm_1$ plus light--quarks and $\tilde{\chi}^\pm_1$ decaying to a W--boson and a $\tilde{\chi}^0_1$.

\item {\tt stop} corresponds to the production of a pair of stops ($\tilde{t}$) decaying to a top--quark and a $\tilde{\chi}^0_1$.

\item {\tt stopjets} corresponds to pair production of $\tilde{t}$ which can decay on--shell to a $\tilde{\chi}^0_1$ plus a top--quark or to the lightest chargino $\tilde{\chi}^\pm_1$ plus a b--quark. In the last case the $\tilde{\chi}^\pm_1$ decays to a $\tilde{\chi}^0_1$ plus a W--boson. 
    
\item {\tt sbsb} corresponds to sbottoms ($\tilde{b}$) pair production decaying to a $\tilde{\chi}^0_1$ and a b--quark. 

\item {\tt gmsb$\_$glgl} corresponds to $\tilde{g}$ pair production in a GMSB model in which the $\tilde{g}$ has a long decay chain ending with a $\tilde{\chi}^0_1$ decaying to a gravitino and a photon or a Z--boson.

\item {\tt gmsb$\_$neutneut} corresponds to $\tilde{\chi}^0_1$ pair production in a GMSB model in which the $\tilde{\chi}^0_1$ decays to a gravitino and a photon or a Z--boson.

\item {\tt slsl} corresponds to slepton ($\tilde{l}$) pair production with a two--step decay to either the second lightest neutralino ($\tilde{\chi}^0_2$) and to 
a lepton or to a $\tilde{\chi}^\pm_1$ and a neutrino. The $\tilde{\chi}^0_2$ decays to a $\tilde{\chi}^0_1$ and a neutrino and the $\tilde{\chi}^\pm_1$ to a $\tilde{\chi}^0_1$ and a lepton.

\item {\tt snusnu} corresponds to the production of a pair of sneutrinos ($\tilde{\nu}$) of the first and second generation decaying to neutrinos and to a $\tilde{\chi}^0_1$.
    
\item {\tt neutineuti}  corresponds to the production of pairs of neutralinos (i=2, 3 or 4) decaying to lighter either neutralinos or charginos which end up decaying to a $\tilde{\chi}^0_1$ and a Z or W--boson. 

\item {\tt neutineutilep} corresponds to the production of pairs of neutralinos (i=2, 3 or 4) decaying to lighter either neutralinos or charginos which end up decaying to a $\tilde{\chi}^0_1$ and a Z or W--boson which decay leptonically.  

\item {\tt chachalep} corresponds to the production of pairs of $\tilde{\chi}^\pm_1$ decaying to a  $\tilde{\chi}^0_1$ plus a W--boson.

\item {\tt chaneut} corresponds to the production of pairs of $\tilde{\chi}^\pm_1$ and $\tilde{\chi}^0_2$ in which $\tilde{\chi}^\pm_1$ decays to a W--boson and a $\tilde{\chi}^0_1$ and 
$\tilde{\chi}^0_2$ to a Z--boson and to a $\tilde{\chi}^0_1$.
    
\item {\tt chaneutlep} corresponds to production of $\tilde{\chi}^0_1$ and a $\tilde{\chi}^0_3$ and the $\tilde{\chi}^0_1$ decays to a W--boson and a $\chi_0$ and $\tilde{\chi}^0_3$ decays to a $\tilde{\chi}^0_2$ and a pair of leptons. Finally $\tilde{\chi}^0_2$ decays to a $\tilde{\chi}^0_1$ and a pair of leptons. 
    
\item {\tt gllp} corresponds to $\tilde{g}\tilde{g}$ production followed by a two--step decay chain in which $\tilde{g}$ decays to a $\tilde{\chi}^0_1$ plus a pair of light quarks and the  $\tilde{\chi}^0_1$ decays to a lepton plus two light quarks. The last decay is through a RPV $\lambda'$ coupling. 
   
\item {\tt gllpp} corresponds to $\tilde{g}\tilde{g}$ production decaying through a RPV $\lambda'$ coupling to a top--quark, a u--quark and a c--quark.
    
\item {\tt stlp} corresponds to $\tilde{t}\tilde{t}$ production decaying to a b--quark and a lepton through a 
$\lambda'$ RPV coupling.

\item {\tt stlpp} corresponds to $\tilde{t}\tilde{t}$ production decaying to a top--quark and to a $\tilde{\chi}^0_1$ which decays to a top--quark plus a b--quark and a light quark. There is a RPV $\lambda'$ coupling involved in the $\tilde{\chi}^0_1$ decay.

\item {\tt sll} corresponds to $\tilde{l}\tilde{l}$ production decaying to a lepton and a neutrino through a $\lambda$ RPV coupling.

\item {\tt sllp} corresponds to $\tilde{l}\tilde{l}$ production decaying to a pair of light quarks through a $\lambda'$ RPV coupling.

\item {\tt snul} sneutrino pair production of $\tilde{\nu}$ decaying, through a RPV violation $\lambda$ coupling, to a pair of leptons. 

\item {\tt snulp} corresponds to a pair production of $\tilde{\nu}$ decaying through a RPV violation $\lambda'$ coupling to a pair of light quarks. 

\end{enumerate}

\subsection{Test Data: DarkMachines Hackathon Data and Secret Data}

To estimate the sensitivity of the mixture--of--theories training approach and compare it with single--model supervised and unsupervised methods, we used the same test datasets as in Ref. \cite{Aarrestad:2021oeb}.
These data consist of a SM prediction with all relevant SM background processes for all channels and two datasets with signals.
The SM dataset, consists of $\sim$ 1 billion simulated LHC events corresponding to $10~\rm{fb}^{-1}$ of proton--proton collisions at a center--of--mass energy of 13 TeV. 

We refer to the first signal dataset as the {\tt hackathon data} and the second signal dataset as the {\tt secret data}. For the {\tt hackathon data}, we know what signals are in the dataset, while we know nothing about the signal content of the {\tt secret data}.
The first dataset helped us determine the performance of our approaches signal--by--signal and compare them with other methods in Ref.~\cite{Aarrestad:2021oeb}.
The second dataset provides a second unbiased comparison and is used to see how robust (given different signal assumptions) our conclusions from the first dataset ultimately are.

Since the {\tt secret data} is an unlabelled dataset, we sent the results of all SSA and MoT classifiers to the person in possession of the true labels to finally evaluate the signal improvements on the {\tt secret data} \footnote{The true labels are  still unknown to us.}. Finally, we compared our results with the unsupervised methods studied in Ref.\cite{Aarrestad:2021oeb}.

\section{Building and Training the Classifiers} 
\label{sec:supervised}

\subsection{Figures of Merit}
To evaluate the performance of the different approaches we used two different metrics. 
The first performance metric we used is the area under curve (AUC), which is commonly used for classification problems. The AUC is defined as the area under the 
receiver operating characteristic (ROC) curve. When combining the AUC of many BSM signal models we compute a ``median AUC''.
The other performance metrics are based  on the significance improvement (SI) used in Ref.\cite{Aarrestad:2021oeb}. The significance is defined as usual:
\begin{equation}
\sigma_S\equiv\frac{S}{\sqrt{B}} .
\end{equation}
with S the number of signal events and B being the number of background events after a channel selection requirement.
From Eq. (4.1) one defines the significance improvement (SI) as
\begin{equation}
\sigma_S^\prime=\frac{S'}{\sqrt{B'}}=\frac{\epsilon_S S}{\sqrt{\epsilon_BB}}=\frac{\epsilon_S}{\sqrt{\epsilon_B}}\sigma_S \quad \Rightarrow \quad {\rm SI}\equiv\frac{\epsilon_S}{\sqrt{\epsilon_B}},
\end{equation}
where $\epsilon_S$ and $\epsilon_B$ are selection efficiencies equaling to $S'/S$ and $B'/B$ respectively. $S'$ is the number of signal events after a selection cut on a classifier and $B'$ the number of background events after that cut. 
For different selections of $\epsilon_B$, the values of SI are different. We calculated $\epsilon_S$ at three different background efficiencies $\epsilon_B=10^{-2},\,10^{-3},\,10^{-4}$ for comparison with the results of the {\tt DarkMachines Challenge} and then selected the largest SI as the maximum significance improvement (mSI). Beyond the mSI, the total improvement (TI) is defined as the largest mSI when a signal appears in multiple channels in the test set.

In this section, all channels are considered individually. Therefore, mSI is always equivalent to TI. In the next section, ML models from the four channels are combined for comparison, and therefore TI is different from mSI. We have used TI (mSI, for single channel figures) as the second performance measure.

\subsection{Machine Learning Classifier}
In this work, we used the  Light Gradient Boosting Machine (LightGBM\footnote{We use {\tt LightGBM} (github.com/microsoft/LightGBM) with {\tt binary cross entropy} as loss function, {\tt auc} as early stop metric, 5000 estimators, 500 leaves, 0.01 learning rate, {\tt gbdt} boost type, and max depth equaling to 15 for this study.}, here called BDT)  and a fully connected deep neural network with only linear layers (fcNN) using {\tt binary cross entropy} as loss function and {\tt auc} as early stop metric. For the fcNN, we optimised the number of layers, number of nodes per layer, dropout rates, learning rate, etc., while for the BDT we optimised the number of leaves, maximum depth, number of estimators, learning rate, etc. Since the performance of both was comparable and training a BDT on both GPU and CPU is usually much faster (i.e. $\sim$ 5 times), we used the BDT for building all ML models. No attempt was made to optimise the architecture of the classification network, as this is outside the scope of this work. Improvements obtained by better classifications are likely to be in addition to the improvements obtained by different training methods.

As for the data coding, for each event only the $E^{miss}_T$, $\phi_{E^{miss}_T}$, the number of each object type and the $E$, $p_T$, $\eta$ and $\phi$ of the leading seven jets, the four leading b--jets, the two leading electrons, the two leading positrons, the two leading muons and the two leading anti--muons were used as input to the BDT and fcNN models. If the number of objects per type is less than the required number, zero padding has been applied.

\subsection{Training Procedure and Training Data Size Dependence}
In the following, we describe the details of the training procedure for MoT and SSA. We also compare the dependence of the classifiers on the size and admixtures of the training data.

\subsubsection{Training Data for the Mixture of Theories (MoT) Training} 
\label{MoT}

For the MoT training scheme, we used several BSM signals with 10K events per signal. Following the procedure described in chapter 2 for the four different analysis channels, we used (37, 51, 18, 71) BSM signals for (channel 1, channel 2a, channel 2b, channel 3) respectively. 

In this section, we want to test the dependence of all our classifiers on the size of the training data (number of events in the training dataset) and on the composition of the mixture of theories. Therefore, instead of directly using the whole training dataset, we gradually add more signals to the training dataset in steps of (5, 7, 2, 10) for the 4 channels\footnote{(5, 7, 2, 10) signals correspond to (50K, 70K, 20K, 100K) events}. The different steps were chosen based on the different total number of BSM models available per channel\footnote{Specifically for channel 1, we trained ML models  on (1, 6, 11, 16, 21, 26, 31, 36, 37) signals. For channel 2a, we trained ML models on (1, 8, 15, 22, 29, 36, 43, 50, 51) signals. For channel 2b, we trained ML models on (1, 3, 5, 7, 9, 11, 13, 15, 17, 18) signals and for channel 3, we trained ML models on (1, 11, 21, 31, 41, 51, 61, 71) signals.}.
For each classifier, the order of ML signals is randomized before adding them to the training dataset, step by step. This process is repeated five times, resulting in five different randomizations. In total, we trained $(9 + 9 + 10 + 8) * 5 = 180$ ML models. With this procedure, we can test whether the order or the model admixture has an impact on the performance.
Finally, we have 5 different ML models for each channel with the full training set, but each of the 5 ML models has a different randomization of the training data.

We are mainly interested in the question at how many BSM models the accuracy of the MoT classifiers reaches saturation, i.e. no longer improves. The results are summarised in Fig.~\ref{fig:AUC1}. The figure shows that for about 10 signals, the MoT classifiers for channel 1 and channel 3 saturate. Fig.~\ref{fig:AUC2} in the Appendix~\ref{app:EF} shows similar behaviour for channels 2a and 2b. However, the small median AUC in channel 2a and the smaller set of training data and models lead to a less clear conclusion for these channels. For more details, see Appendix~\ref{app:EF} (Figs.~\ref{fig:AUC2}, \ref{fig:mSI1}, \ref{fig:mSI2}).

\subsubsection{Single Signal Approach (SSA) Training}
\label{SSA}
To understand the impact of the size of the training data, we trained many SSA classifiers for different BSM signals and -- for each signal -- in steps of 10K events. It is important to note that 100K events in the following comparisons represent 10 signals for the MoT scheme with 10k events per signal, while 100k events represent 1 signal with 100k events for the SSA scheme. In total, we have 360 SSA classifiers for 4 channels. The details of the signal we used to train SSA classifiers are described in the Appendix~\ref{app:SSAC}.

In Fig.~\ref{fig:AUC1} we can see that SSA classifiers perform quite differently from MoT classifiers and generally have a lower AUC. It can also be seen that the lower AUC is not due to the size of the training data.

\begin{figure}[!htbp]
    \center
    \includegraphics[width=0.8\textwidth]{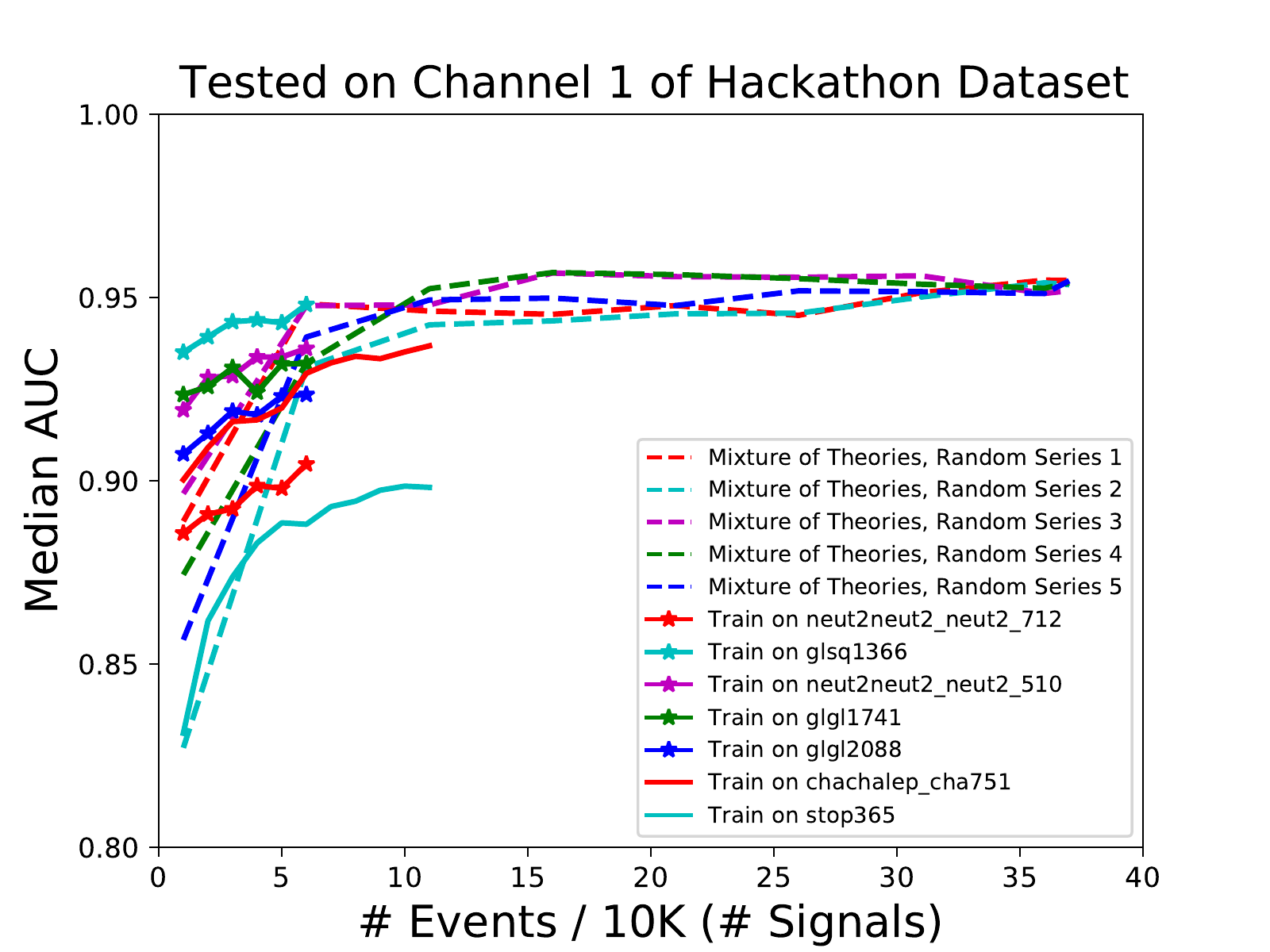}
    \includegraphics[width=0.8\textwidth]{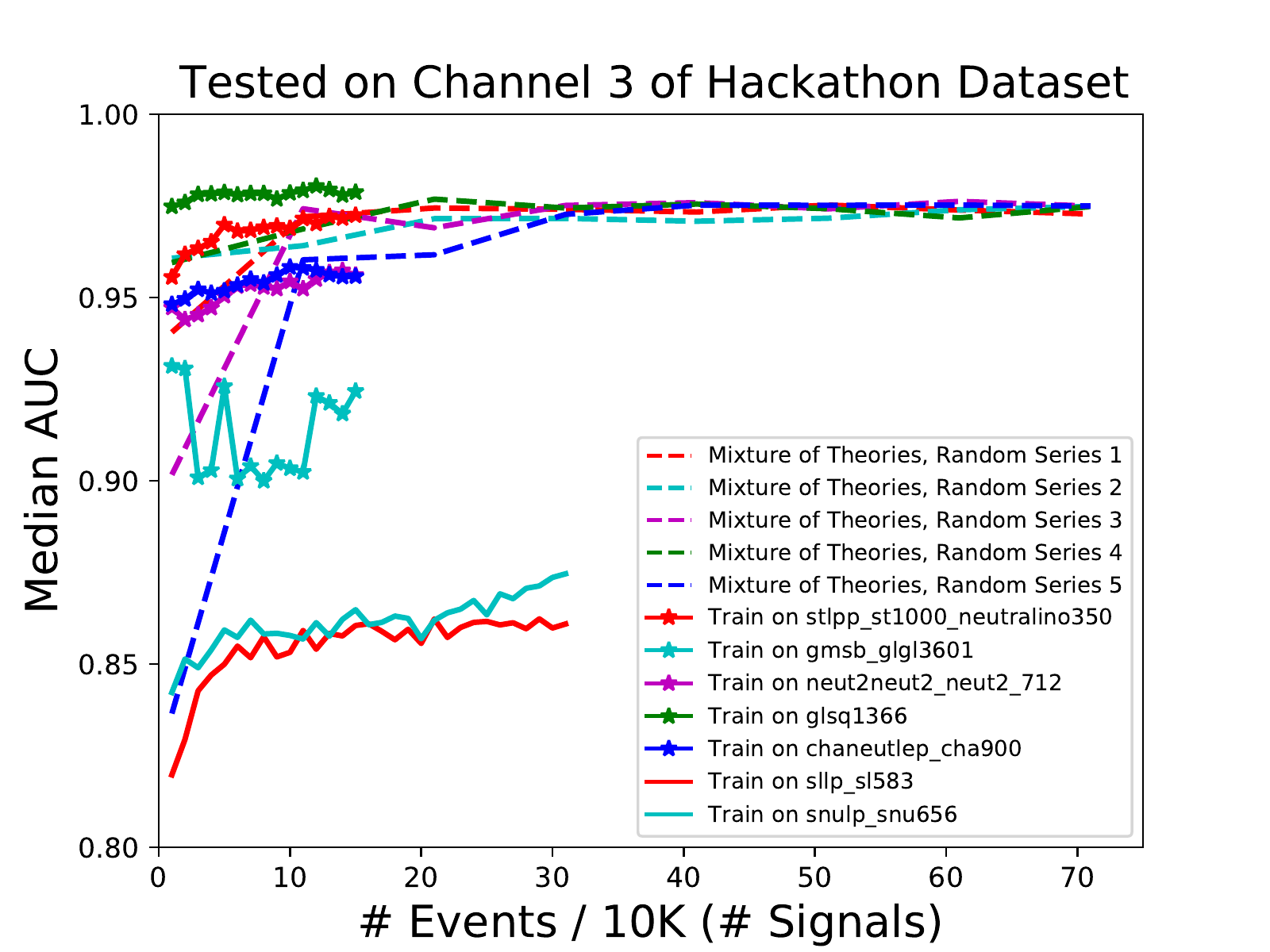}
    \caption{The figure shows the median AUCs of the trained ML models for channels with hadronic final states, tested on signals from the {\tt hackathon data}. For the dashed line, the training processes based on an increasing number of signals with a point size of 10k are repeated 5 times per channel in random rows. The steps of adding signals to the training set are (5, 10) for (channel 1, channel 3), while all channels with the ML model start with only one signal in the training set. For the star--shaped lines, we use only 1 single signal and gradually increase the number of events in steps of 10k. The solid lines are similar to the star lines trained with a single signal. However, additional events are generated to extend the line.}
\label{fig:AUC1}
\end{figure}

\subsection{Performance for Different BSM Test Signals}
\label{MoTvsSSA}

\begin{figure}[!htbp]
    \center
    \includegraphics[width=\textwidth]{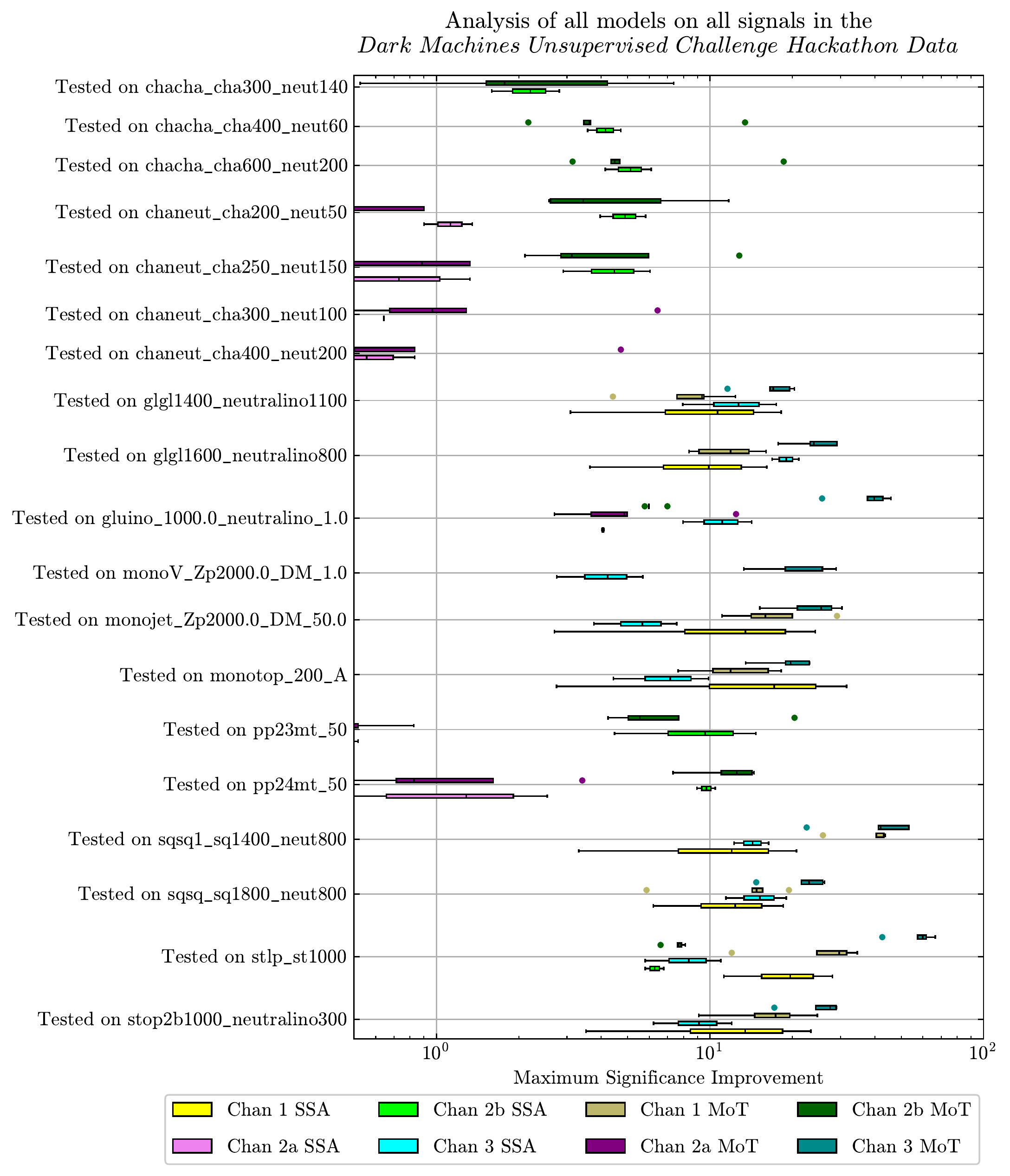}
    \caption{The figure shows the results for the maximum significance improvements tested with signals from the {\tt hackathon data}. This compares the supervised methods trained with a mixture of theories (boxes above the middle lines) with those trained with a single signal approach (boxes below the middle lines). Chan X SSA means the results for channel X from the single signal approach. Chan X MoT means the results for channel X from a mixture of theories. SSA contains 2 ML models corresponding to the endpoints of the solid lines in Figs.~\ref{fig:AUC1}, \ref{fig:AUC2}, \ref{fig:mSI1} and \ref{fig:mSI2}, while MoT contains 5 ML models corresponding to the points on the dashed lines that have the same event number as the endpoints of the solid lines.}
\label{fig:mSI3}
\end{figure}

In this section, we measure the performance of our supervised classifiers on different BSM test signals. For this purpose, we plot the maximum significance improvement (mSI)\footnote{We took two SSAs and five MoTs trained on the same number of events to compare these different schemes. For channel 1 we used 110k events for training, for channel 2a 150k, for channel 2b 180k and finally for channel 3 310k events. Our figures are in the same style as those in Ref. ~\cite{Aarrestad:2021oeb}. Essentially, a coloured box is drawn that spans the inner half of the data. A black line through the box marks the median. Whiskers extend from the box to either the maximum or minimum, provided they are no further than 1.5 box lengths from the edge of the box. The remaining outlier points are shown as coloured circles.} in Fig.~\ref{fig:mSI3}, where the different channels of the supervised methods are shown in different colours. Other quantities such as AUC and signal efficiencies can be found in the Appendix~\ref{app:EF} (Fig.~\ref{fig:AUC3}). 

Fig.~\ref{fig:mSI3} shows that, in terms of mSI, there are SSAs which have comparable performance to MoT for some specific signals. However, others are much worse. This is because, as expected, the test signal leads to a different collider signature than the signals used for training. For the same test signal, one SSA classifier may perform well because it has been trained on data with a similar collider signature, while the other SSA performs very poorly because it has been trained on data with a very different signature. Therefore, the variance for the SSA classifier can be very large for most test signals in Fig.~\ref{fig:mSI3}.

In summary, we find that MoT is generally much better than SSA, both in terms of stability and performance, especially for unknown test signals that exclude the single signal specifically selected for training.

\section{Comparison with Unsupervised Methods}
\label{sec:unsupervised}

In this section, we compare our results with the unsupervised methods used in Ref. ~\cite{Aarrestad:2021oeb} in two ways: with {\tt hacktathon test data} and with {\tt secret data}. While in Ref.~\cite{Aarrestad:2021oeb} classifiers were compared with others on all 4 channels, however, our ML models were trained on only one channel. Therefore, we combined 4 classifiers, one for each channel, into a new classifier. For comparison, we construct 7 MoT classifiers trained on all signals for 4 channels, for a total of $37 + 51 + 18 + 71 = 177$ signals. In addition, we construct $7^4=2401$ SSA classifiers trained on 4 signals, one per channel. Finally, to compare MoT with SSA on the same data set, we construct another 7 mini--MoT classifiers trained on randomly selected signals containing the same event number as the training set of the SSA scheme \footnote{We have combined the first MoT models for each channel into a new MoT model that includes all channels. The new MoT model for 4 channels was created from 4 MoT models that were trained individually for only one channel. The other 4 MoT models for all channels were created in the same way. In contrast, all combinations of 7 SSA models for 4 channels, i.e. 2401 in total, were considered.}.

\subsection{Comparison on Hackathon Data}

In this section we compare the performance of the supervised models with that of the unsupervised models using the {\tt hackathon data}, containing a mixture of SUSY and non--SUSY signals. Fig. 3 shows a direct comparison of the overall improvements for both approaches. While the supervised methods perform better than the unsupervised methods for Min TI and Median TI, this is not quite true for Max TI \footnote{It is important to highlight that supervised methods use a BDT or simple network architectures compared to some unsupervised models.}. 
We see that some unsupervised models based on the combination of a Flow model and a set of Deep SVDD models (see Ref.~\cite{Aarrestad:2021oeb} for details on these models) perform better. The reason why the overall performance of the supervised methods on the {\tt hackathon data} is good is because we construct a training dataset that has sufficiently similar collider signatures for the classifiers to learn the difference between SM and BSM signatures in the {\tt hackathon data}.

It is interesting to note that the saturation effect is visible when comparing the Mini MoT with the MoT for all the quantities and the results are mostly not sensitive to the order of the events in the admixture of theory models approach.  

The details of the comparison for each test signal can be found in the Appendix~\ref{app:EF} (Figs.~\ref{fig:AUC4} and~\ref{fig:mSI4}).

\begin{figure}[!htbp]
    \center
    \includegraphics[width=\textwidth]{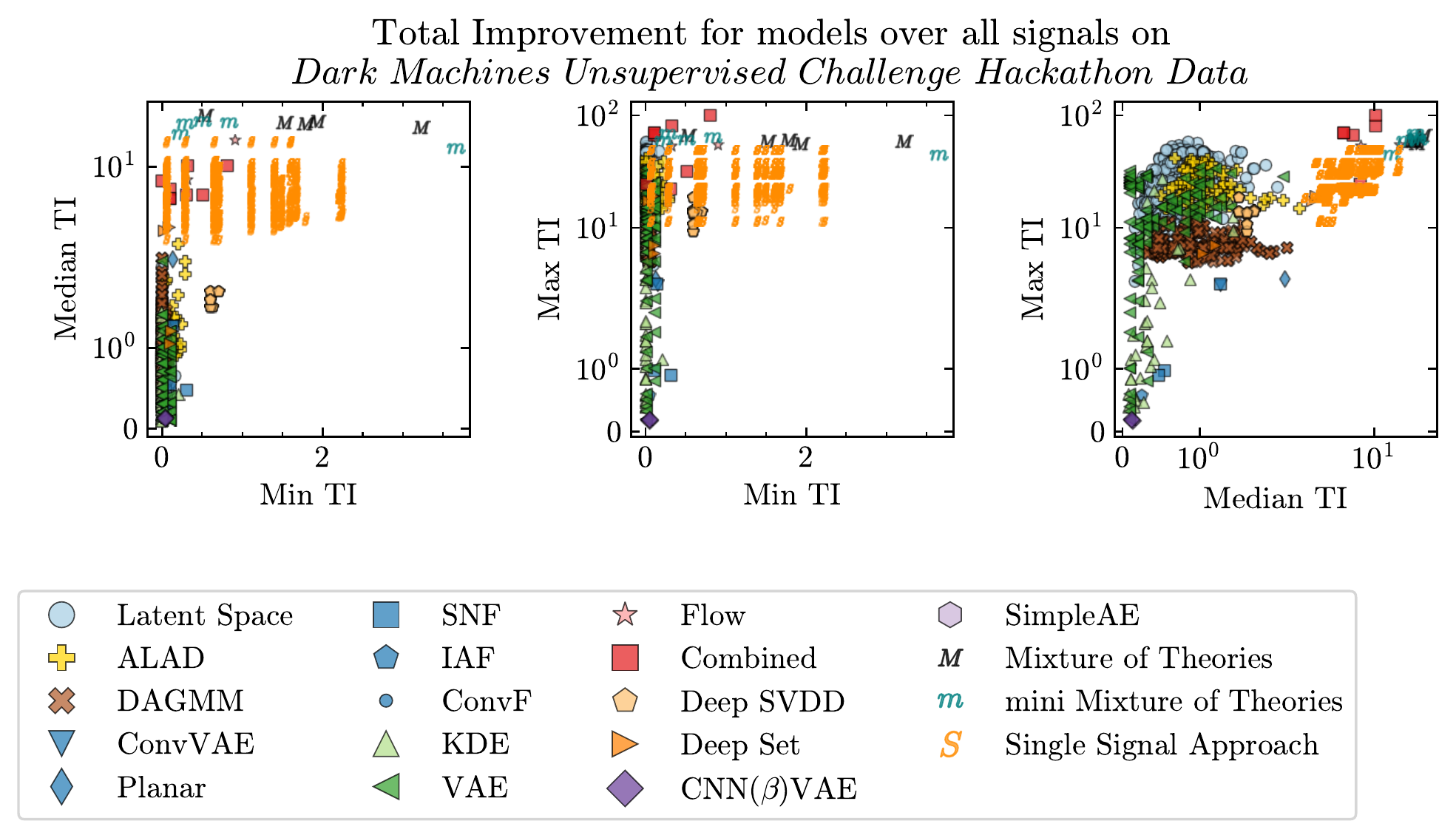}
    \caption{The figure shows the results for the total improvements of {\tt DarkMachines hackathon data} comparing the supervised methods and the unsupervised methods. There are a total of $2401$ SSA models denoted as $S$, 5 MoT models trained on all signals denoted as $M$, and 5 mini MoT models trained on a randomly selected subset of the signals denoted as $m$.}
\label{fig:TI3}
\end{figure}

\subsection{Comparison on Secret Data}

\begin{figure}[htb]
    \center
    \includegraphics[width=\textwidth]{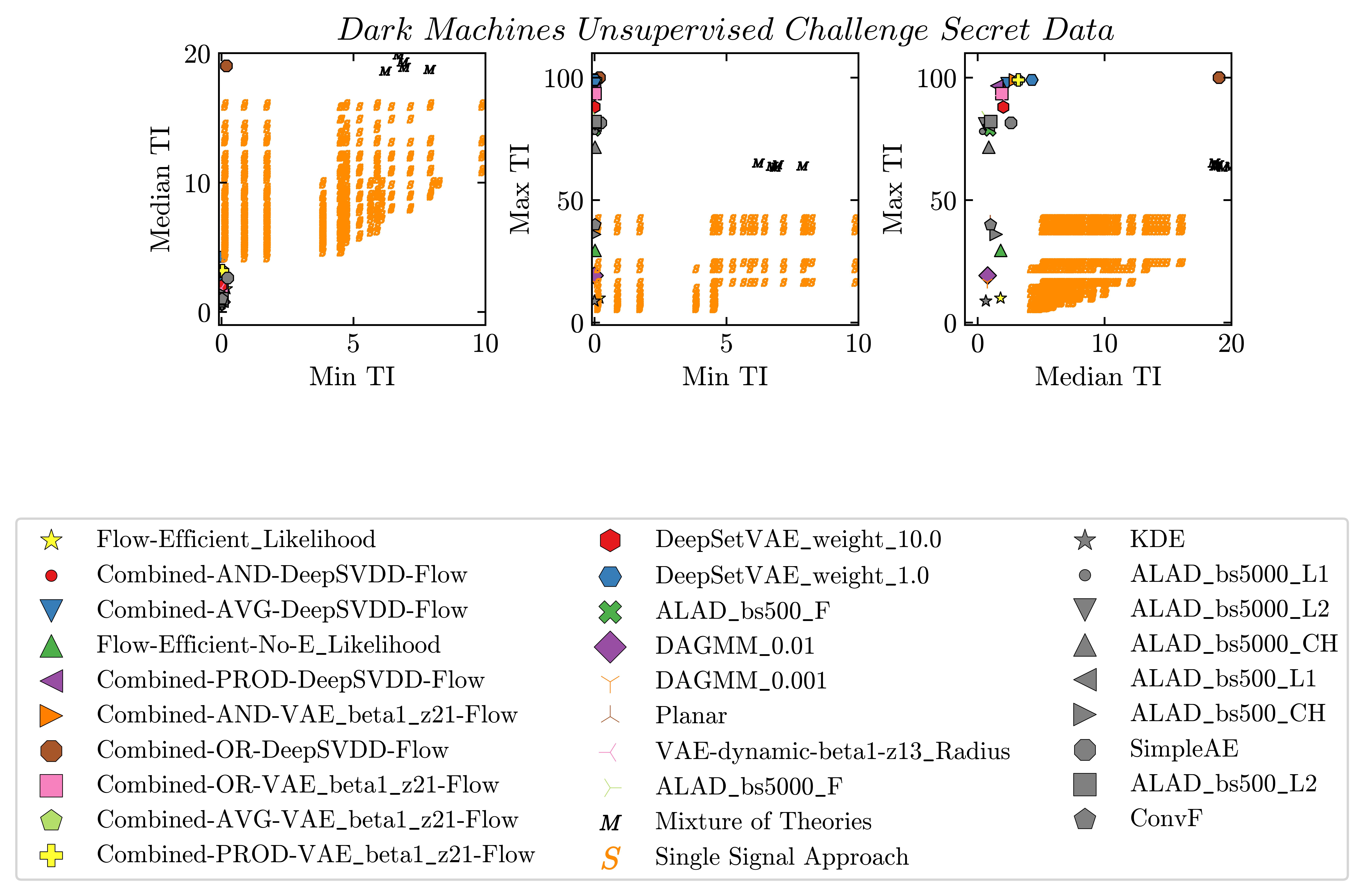}
     \caption{The figure shows the results for the total improvements of the {\tt secret data} comparing the supervised and unsupervised methods. There are a total of $2401$ SSA models denoted as $S$, and 5 MoT models trained on all signals denoted as $M$.}
\label{fig:TI4}
\end{figure}

In this section we repeat the comparison made above, but with the {\tt secret data}, which allows for an unbiased comparison since we do not know the signals contained in this dataset. Fig.~\ref{fig:TI4} shows the results of the comparison of the supervised and unsupervised models for the total improvements. In this case, we see that Min TI and Median TI perform much better than the unsupervised models, although now a large number of unsupervised models significantly outperform the supervised ones. The reason for this is that only the best performing unsupervised models are applied on the secret data and shown in this plot (for details see Ref. ~\cite{Aarrestad:2021oeb}). 
It is interesting to see that the best unsupervised approaches outperform all supervised approaches.
Finally, MoT's good performance on completely unknown signals proves its potential ability to detect new physics in LHC data.

\section{Conclusions}
\label{sec:results}

In this study, we compare supervised training strategies using a single BSM model approach with unsupervised approaches and a new training scheme called Mixture of Theories (MoT). We propose MoT as a strategy in which a signaling hypothesis is constructed from a commingled set of many different theoretical hypothetical signaling models.
In general, we find that classifiers built with the MoT approach and trained with a mix of 10 or more different BSM models (and parameter sets) perform very well.
They typically outperform ML models trained with individual BSM models in our comparisons with test signals.

Two other data sets are considered in this study for comparison, namely the {\tt  hackathon data} and the {\tt secret data} from a previous comparison of anomaly detection algorithms with LHC data. We found that the supervised single--model strategies can outperform many unsupervised strategies in the {\tt hackathon data}. However, the MoT models again outperform the supervised single model strategies. Finally, when applied to the {\tt secret data}, the MoT strategies outperform all but the best unsupervised strategies.

We conclude that the MoT strategy offers an alternative supervised training approach for searching for new physics in a less model-dependent way. We propose to add signal regions to LHC searches based on the best unsupervised strategies and the MoT strategies.

\FloatBarrier

\Acknowledgements 

The author(s) gratefully acknowledges the computer resources at Artemisa, funded by the European Union ERDF and Comunitat Valenciana as well as the technical support provided by the Instituto de Fisica Corpuscular, IFIC (CSIC-UV). R. RdA also  acknowledges the Ministerio de Ciencia e Innovación (PID2020-113644GB-I00).

The authors thank Melissa van Beekveld for providing the performance numbers for the secret dark machines dataset.

\appendix
\setcounter{equation}{0}
\label{app}
\APPENDIX
\section{Details of Training Dataset}
\label{app:data}

In this section, we explain how we name each signal we used for training. You can find the files with the same names given in this section online for further investigation. In the file names, the masses of the relevant particles are given in GeV.

\subsection{Simplified Models}

{\tt btzpLoop}: In the signal name {\tt btzpLoopAX/VX} means the $m_{Z^\prime}=X$ GeV, while the couplings between mediator and quarks are axial--vector/vector like. After event selection for the four channels,

\begin{itemize}
    \item Channel 1: 
    
    {\tt btzpLoopA5} and {\tt btzpLoopA5} 
    \item Channel 3:
    
    {\tt btzpLoopA50}, {\tt btzpLoopA40}, 
    {\tt btzpLoopA30}, {\tt btzpLoopA20}, 
    
    {\tt btzpLoopA10}, {\tt btzpLoopA5}, 
    {\tt btzpLoopV50}, {\tt btzpLoopV40}, 
    
    {\tt btzpLoopV30}, {\tt btzpLoopV20}, 
    {\tt btzpLoopV10}, {\tt btzpLoopV5} 
\end{itemize}

{\tt LReff}: In the signal name {\tt LReff\_X\_Y}, X is $m_{W^\prime}$, while Y is $m_{Z^\prime}$. After event selection for the four channels,

\begin{itemize}
    \item Channel 2a: 
    
    {\tt LReff\_100\_10}, {\tt LReff\_100\_100}, {\tt LReff\_100\_500}, 
    
    {\tt LReff\_500\_10}, {\tt LReff\_500\_100}, {\tt LReff\_500\_500}
    \item Channel 3: 
    
    {\tt LReff\_10\_10}, {\tt LReff\_50\_10},
    {\tt LReff\_100\_10}, {\tt LReff\_500\_10} 
\end{itemize}

{\tt type2seesaw}: In the signal name {\tt type2seesaw\_X\_Y}, X is the mass of extra Higgs, while Y is the masses of new charged scalars. After event selection for the four channels, 

\begin{itemize}
    \item Channel 2a: 
    
    {\tt type2seesaw\_100\_150}, {\tt type2seesaw\_150\_150},
    {\tt type2seesaw\_150\_200}, 
    
    {\tt type2seesaw\_200\_200}, {\tt type2seesaw\_200\_250}, 
    {\tt type2seesaw\_250\_250}, 
    
    {\tt type2seesaw\_250\_300}, {\tt type2seesaw\_300\_300}, 
    {\tt type2seesaw\_350\_350}, 
    
    {\tt type2seesaw\_400\_400}, {\tt type2seesaw\_450\_450}, 
    {\tt type2seesaw\_500\_500} 
    \item Channel 2b: 
    
    {\tt type2seesaw\_150\_150}, {\tt type2seesaw\_200\_200}, 
    {\tt type2seesaw\_250\_250}, 
    
    {\tt type2seesaw\_300\_300}, {\tt type2seesaw\_350\_350},
    {\tt type2seesaw\_400\_400},
    
    {\tt type2seesaw\_450\_450}, {\tt type2seesaw\_500\_500}, 
    \item Channel 3:
    
    {\tt type2seesaw\_500\_500}
\end{itemize}

{\tt Wprime}: In the signal name {\tt wprimeAX/VX}, X is $m_{W^\prime}$, while V (A) means vector (axial--vector) coupling. After event selection for the four channels, 

\begin{itemize}
    \item Channel 2a:

    {\tt wprimeA10}, {\tt wprimeA20}, {\tt wprimeA100}, 
    {\tt wprimeA200}, {\tt wprimeA300}, 
    
    {\tt wprimeA400}, {\tt wprimeA500}, {\tt wprimeV10}, 
    {\tt wprimeV20}, {\tt wprimeV50},
    
    {\tt wprimeV100}, {\tt wprimeV200},
    {\tt wprimeV300}, {\tt wprimeV500}
\end{itemize}

{\tt mDM}: In the signal name {\tt mDMX}, X means $m_S$. After event selection for 4 channels, only {\tt mDM500} appears in channel 3.

\subsection{SUSY Models}

{\tt HH}: In the signal name {\tt HH\_X}, X is $m_{H^{\pm}}$. After event selection for the four channels,

\begin{itemize}
    \item Channel 1:
    
     {\tt HH\_{103}}, {\tt HH\_754} 
      
    \item Channel 2a:
    
     {\tt HH\_150}
     
    \item Channel 3:
    
     {\tt HH\_{103}}, {\tt HH\_150}, {\tt HH\_754}
\end{itemize}    

{\tt hh}: In the signal name {\tt hh\_X}, X is $m_H$. After event selection for the four channels,

\begin{itemize}
    \item Channel 2a:
    
     {\tt hh\_1000}
     
    \item Channel 2b:
    
     {\tt hh\_1000}
     
    \item Channel 3:
    
     {\tt hh\_100}
\end{itemize} 

{\tt AA}: In the signal name {\tt AA\_X}, X is $m_A$. After event selection for the four channels,

\begin{itemize}
    \item Channel 3:
    
     {\tt AA\_600}, {\tt AA\_924} 
\end{itemize} 

{\tt glgl}: In the signal name {\tt glglX}, X is $m_{\tilde{g}}$. 
After event selection for the four channels

\begin{itemize}
    \item Channel 1:
    
     {\tt glgl\_1178} with $m_{\tilde{\chi}_1^0} \sim 100$ GeV, {\tt glgl\_1680} with $m_{\tilde{\chi}_1^0} \sim 50$ GeV, {\tt glgl\_1741} with $m_{\tilde{\chi}_1^0} \sim 200$ GeV, {\tt glgl\_1818} with $m_{\tilde{\chi}_1^0} \sim 100$ GeV, 
     {\tt glgl\_2088} with $m_{\tilde{\chi}_1^0} \sim 500$ GeV  
     
    \item Channel 3:

    Same as in Channel 1
\end{itemize}  

{\tt sqsq}: In the signal name {\tt sqsq\_sqX}, X is $m_{\tilde{q}}$. 
After event selection for the four channels

\begin{itemize}
    \item Channel 1:
    
     {\tt sqsq\_sq766} with $m_{\tilde{\chi}_1^0} \sim 249$ GeV
     
    \item Channel 3:

    Same as in Channel 1
\end{itemize}  

{\tt glsq}: In the signal name {\tt glsqX\_Y}, X is $m_{\tilde{g}}$ and Y is $m_{\tilde{q}}$. 
After event selection for the four channels

\begin{itemize}
    \item Channel 1:
    
     {\tt sqsq1366\_1562} with $m_{\tilde{\chi}_1^0} \sim 299$ GeV
     
    \item Channel 3:

    Same as in Channel 1
\end{itemize}  

{\tt stop}: In the signal name {\tt stopX}, X is $m_{\tilde{t}}$. 
After event selection for the four channels

\begin{itemize}
    \item Channel 1:
    
     {\tt stop365} with $m_{\tilde{\chi}_1^0} = 50$ GeV, {\tt stop846} with $m_{\tilde{\chi}_1^0} \sim 500$ GeV
     
\end{itemize}  

{\tt stopjets}: In the signal name {\tt stopjetsX}, X is $m_{\tilde{t}}$. 
After event selection for the four channels

\begin{itemize}
    \item Channel 1:
    
     {\tt stopjets845} with $m_{\tilde{\chi}_1^0} = 249$ GeV
     
    \item Channel 3:

     Same as in Channel 1
     
\end{itemize}  

{\tt sbottom}: In the signal name {\tt sbottomX}, X is $m_{\tilde{b}}$. 
After event selection for the four channels

\begin{itemize}
    \item Channel 1:
    
     {\tt sbottom198} with $m_{\tilde{\chi}_1^0} = 100$ GeV, {\tt sbottom681} with $m_{\tilde{\chi}_1^0} \sim 300$ GeV, {\tt sbottom781} with $m_{\tilde{\chi}_1^0} \sim 326$ GeV 

   \item Channel 3:

     Same as in Channel 1
     
\end{itemize} 

{\tt gmsb\_glgl}: In the signal name {\tt gmsb\_glglX}, X is $m_{\tilde{g}}$. 
After event selection for the four channels

\begin{itemize}
    \item Channel 1:
    
     {\tt gmsb\_glgl3601} with $m_{\tilde{\chi}_1^0} = 721$ GeV
     
   \item Channel 2a:

     Same as in Channel 1     
     
   \item Channel 3:

     Same as in Channel 1
     
\end{itemize} 

{\tt gmsb\_neutneut}: In the signal name {\tt gmsb\_neuneut\_neutX}, X is $m_{\tilde{\chi}_1^0}$
After event selection for the four channels

\begin{itemize}
    \item Channel 1:
    
     {\tt gmsb\_neutneut\_neut721} with  $m_{\tilde{\chi}_1^0} = 721$ GeV
    
   \item Channel 3:

     Same as in Channel 1
     
\end{itemize} 

{\tt slsl}: In the signal name {\tt slsl\_slepX}, X is $m_{\tilde{l}}$.
After event selection for the four channels

\begin{itemize}
    \item Channel 2a:
    
     {\tt slsl\_slep700} with  $m_{\tilde{\chi}_1^0} = 399$ GeV
    
   \item Channel 2b:

     Same as in Channel 2a
     
    \item Channel 3:

      {\tt slsl\_slep700} with  $m_{\tilde{\chi}_1^0} = 399$ GeV, {\tt slsl\_slep1006} with  $m_{\tilde{\chi}_1^0} = 426$ GeV

\end{itemize} 

{\tt neutineuti}: In the signal name {\tt neutineuti\_neutiX}, X is $m_{\tilde{\chi}_i^0}$.
After event selection for the four channels

\begin{itemize}

    \item channel 1:
    
      {\tt neut2neut2\_neut2\_712} with $m_{\tilde{\chi}_1^0} = 443$ GeV, {\tt neut4neut4\_neut4\_923} with $m_{\tilde{\chi}_1^0} = 444$ GeV
     
    \item Channel 3:
    
      {\tt neut2neut2\_neut2\_510} with $m_{\tilde{\chi}_1^0} = 99$ GeV, {\tt neut4neut4\_neut4\_923} with $m_{\tilde{\chi}_1^0} = 97$ GeV 
      
\end{itemize} 

{\tt neutineutilep}: In the signal name {\tt neutineutilep\_neutiX}, X is $m_{\tilde{\chi}_i^0}$.
After event selection for the four channels

\begin{itemize}

    \item channel 1:
    
     {\tt neut3neut3lep\_neut3\_616} with $m_{\tilde{\chi}_1^0} = 399$ GeV, {\tt neut4neut4lep\_neut4\_923} with $m_{\tilde{\chi}_1^0} = 97$ GeV
 
    \item channel 2a:
    
      {\tt neut3neut3lep\_neut3\_616} with $m_{\tilde{\chi}_1^0} = 399$ GeV, 
      {\tt neut4neut4lep\_neut4\_923} with $m_{\tilde{\chi}_1^0} = 97$ GeV 

    \item channel 2b:
    
      {\tt neut3neut3lep\_neut3\_616} with $m_{\tilde{\chi}_1^0} = 399$ GeV

    \item channel 3:
    
      {\tt neut3neut3lep\_neut3\_616} with $m_{\tilde{\chi}_1^0} = 399$ GeV, {\tt neut4neut4lep\_neut4\_923} with $m_{\tilde{\chi}_1^0} = 97$ GeV

\end{itemize} 

{\tt chachalep}: In the signal name {\tt chachalep\_chaX}, X is $m_{\tilde{\chi}_1^\pm}$.
After event selection for the four channels

\begin{itemize}

    \item channel 1:
    {\tt chachalep\_cha510} with $m_{\tilde{\chi}_1^0} = 99$ GeV, {\tt chachalep\_cha751} with $m_{\tilde{\chi}_1^0} = 495$ GeV
    
    \item channel 2a:
    {\tt chachalep\_cha510} with $m_{\tilde{\chi}_1^0} = 99$ GeV, {\tt chachalep\_cha751} with $m_{\tilde{\chi}_1^0} = 495$ GeV
    
    \item channel 3:
    {\tt chachalep\_cha510} with $m_{\tilde{\chi}_1^0} = 99$ GeV, {\tt chachalep\_cha751} with $m_{\tilde{\chi}_1^0} = 495$ GeV
    
\end{itemize} 

{\tt chaneuti}: In the signal name {\tt chaneuti\_cha\_X\_neuti\_Y}, X is $m_{\tilde{\chi}_1^\pm}$ and Y is $m_{\tilde{\chi}_i^0}$. After event selection for the four channels

\begin{itemize}

    \item channel 1:
    
      {\tt chaneut3\_cha\_363\_neut3\_711} with $m_{\tilde{\chi}_1^0} = 99$ GeV
     
    \item channel 2a:

      {\tt chaneut3\_cha\_363\_neut3\_711} with $m_{\tilde{\chi}_1^0} = 99$ GeV
     
    \item Channel 3:

      {\tt chaneut\_cha776\_neut762}, {\tt chaneut\_cha867}
      
\end{itemize} 

{\tt chaneutilep}: In the signal name {\tt chaneutilep\_cha\_X\_neuti\_Y}, X is $m_{\tilde{\chi}_1^\pm}$ and Y is $m_{\tilde{\chi}_i^0}$.
After event selection for the four channels

\begin{itemize}

    \item channel 1:

      {\tt chaneutlep\_cha\_900} with $m_{\tilde{\chi}_1^0} = 197$ GeV, {\tt chaneut2lep\_cha700} with $m_{\tilde{\chi}_2^0} = 700$ GeV and $m_{\tilde{\chi}_1^0} = 297$ GeV, {\tt chaneut2lep\_cha900} with $m_{\tilde{\chi}_2^0} = 900$ GeV and $m_{\tilde{\chi}_1^0} = 197$ GeV  

    \item channel 2a:

      {\tt chaneut2lep\_cha700} with $m_{\tilde{\chi}_2^0} = 700$ GeV and $m_{\tilde{\chi}_1^0} = 297$ GeV 
       
    \item channel 2b:
    
      {\tt chaneut2lep\_cha700} with $m_{\tilde{\chi}_2^0} = 700$ GeV and $m_{\tilde{\chi}_1^0} = 297$ GeV, {\tt chaneut2lep\_cha900} with $m_{\tilde{\chi}_2^0} = 900$ GeV and $m_{\tilde{\chi}_1^0} = 197$ GeV, {\tt chaneut3lep\_cha363} with $m_{\tilde{\chi}_3^0} = 711$ GeV $m_{\tilde{\chi}_1^0} = 197$ GeV

    \item Channel 3:

\end{itemize} 

{\tt gllp}: In the signal name {\tt gllp\_glX}, X is $m_{\tilde{g}}$. 
After event selection for the four channels

\begin{itemize}
    \item Channel 3:
    
     {\tt gllp\_gl1800} with $m_{\tilde{\chi}_1^0} = 200$ GeV
     
\end{itemize}  

{\tt gllpp}: In the signal name {\tt gllpp\_lambdapp\_glX}, X is $m_{\tilde{g}}$. 
After event selection for the four channels

\begin{itemize}
    \item Channel 3:
    
     {\tt gllpp\_lambdapp\_gl1500}, {\tt gllpp\_lambdapp\_gl1800} 
     
\end{itemize}  

{\tt stlp}: In the signal name {\tt stlp\_stX}, X is $m_{\tilde{t}}$. 
After event selection for the four channels

\begin{itemize}
    \item Channel 1:
    
     {\tt stlp\_st560}, {\tt stlp\_st1560} 

    \item Channel 2a:
    
     {\tt stlp\_st560}, {\tt stlp\_st1560} 
     
    \item Channel 3:
    
     {\tt stlp\_st560}, {\tt stlp\_st1560}      
     
\end{itemize} 

{\tt stlpp}: In the signal name {\tt stlpp\_stX}, X is $m_{\tilde{t}}$. 
After event selection for the four channels

\begin{itemize}
    \item Channel 1:
    
     {\tt stlpp\_st1000}
     
    \item Channel 3:
    
     {\tt stlp\_st1000}     
     
\end{itemize} 

{\tt sll}: In the signal name {\tt sll\_slX}, X is $m_{\tilde{l}}$. 
After event selection for the four channels

\begin{itemize}
    \item Channel 2a:
    
     {\tt sll\_sl350}
    
    \item Channel 3:
    
     {\tt sll\_sl350}    
    
\end{itemize} 

{\tt sllp}: In the signal name {\tt sllp\_slX}, X is $m_{\tilde{l}}$. 
After event selection for the four channels

\begin{itemize}
    \item Channel 3:
    
     {\tt sllp\_sl583}
    
\end{itemize} 

{\tt snul}: In the signal name {\tt snul\_snuX}, X is $m_{\tilde{\nu}}$. 
After event selection for the four channels

\begin{itemize}
    \item Channel 2a:
    
     {\tt snul\_snu475}
     
    \item Channel 2b:
    
     {\tt snul\_snu475}
    
    \item Channel 3:
    
     {\tt snul\_snu475}    
    
\end{itemize} 

{\tt snulp}: In the signal name {\tt snulp\_snuX}, X is $m_{\tilde{\nu}}$. 
After event selection for the four channels

\begin{itemize}

    \item Channel 3:
    
     {\tt snulp\_snu656}  
     
\end{itemize} 

\section{SSA Classifiers}
\label{app:SSAC}

For all channels, the number of signals passing the event selection is not very large. Therefore, to compare with MoT, we first select the signals with the five largest numbers of events per channel. In doing so, we find that most of the signals with a large number of events passing the selection have heavy BSM particles. Moreover, the number of events is not sufficient to compare with the MoT results. Therefore, we selected two signals with relatively light BSM particles per channel and generated more events for them. The first five signals are called $S1$, $S2$, $S3$, $S4$, $S5$, while the last two are called $S6$ and $S7$. For $S6$ and $S7$, we can compare them with MoT trained on exactly the same event numbers.

\begin{itemize}
    \item For channel 1, 
        \begin{enumerate}
            \item Train on 60K events of {\tt neut2neut2\_neut2\_712} as $S1$\footnote{The details of the notion in the name of the listed signals can be found in Appendix~\ref{app:data}.},
            \item Train on 60K events of {\tt glsq1366} as $S2$,
            \item Train on 60K events of {\tt neut2neut2\_neut2\_510} as $S3$,
            \item Train on 60K events of {\tt glgl1741} as $S4$,
            \item Train on 60K events of {\tt glgl2088} as $S5$.
            \item Train on 110K {\tt chachalep\_cha751} as $S6$
            \item Train on 110K {\tt stop365} as $S7$
        \end{enumerate}

    \item For channel 2a, 
        \begin{enumerate}
            \item Train on 100K events of {\tt neut2neut2lep\_neut2\_900} as $S1$
            \item Train on 100K events of {\tt hh\_1000} as $S2$
            \item Train on 100K events of {\tt chaneut2lep\_cha900} as $S3$
            \item Train on 100K events of {\tt neut3neut3lep\_neut3\_616} as $S4$
            \item Train on 100K events of {\tt slsl\_slep700} as $S5$
            \item Train on 150K {\tt LReff\_100\_100} as $S6$
            \item Train on 150K {\tt type2seesaw\_100\_150} as $S7$
        \end{enumerate}

    \item For channel 2b,
        \begin{enumerate}
            \item Train on 100K events of {\tt neut3neut3lep\_neut3\_616} as $S1$
            \item Train on 100K events of {\tt neut2neut2lep\_neut2\_900} as $S2$
            \item Train on 100K events of {\tt slsl\_slep700} as $S3$
            \item Train on 100K events of {\tt hh\_1000} as $S4$
            \item Train on 100K events of {\tt chaneut3lep\_cha363} as $S5$
            \item Train on 180K {\tt type2seesaw\_150\_150} as $S6$
            \item Train on 180K {\tt type2seesaw\_200\_200} as $S7$
        \end{enumerate}

    \item For channel 3,
        \begin{enumerate}
            \item Train on 160K events of {\tt stlpp\_st1000\_neutralino350} as $S1$
            \item Train on 160K events of {\tt gmsb\_glgl3601} as $S2$
            \item Train on 160K events of {\tt neut2neut2\_neut2\_712} as $S3$
            \item Train on 160K events of {\tt glsq1366} as $S4$
            \item Train on 160K events of {\tt chaneutlep\_cha900} as $S5$
            \item Train on 310K {\tt sllp\_sl583} as $S6$
            \item Train on 310K {\tt snulp\_snu656} as $S7$
        \end{enumerate}
\end{itemize}

For $S1$ to $S5$ of channel 1, we have 60K events to form 6 SSA classifiers with steps of 10K events. For $S6$ and $S7$, we have 110K events to form 11 SSA classifiers. Therefore, we have (6 * 5 + 11 * 2) = 52 SSA classifiers for channel 1. In total, we have trained 
$$(6 * 5 + 11 * 2) + (10 * 5 + 15 * 2) + (10 * 5 + 18 * 2) + (16 * 5 + 31 * 2) = 360\ {\rm ML\ Models}.$$ 

\section{Extra Figures}
\label{app:EF}

\begin{figure}[!htbp]
    \center
    \includegraphics[width=0.8\textwidth]{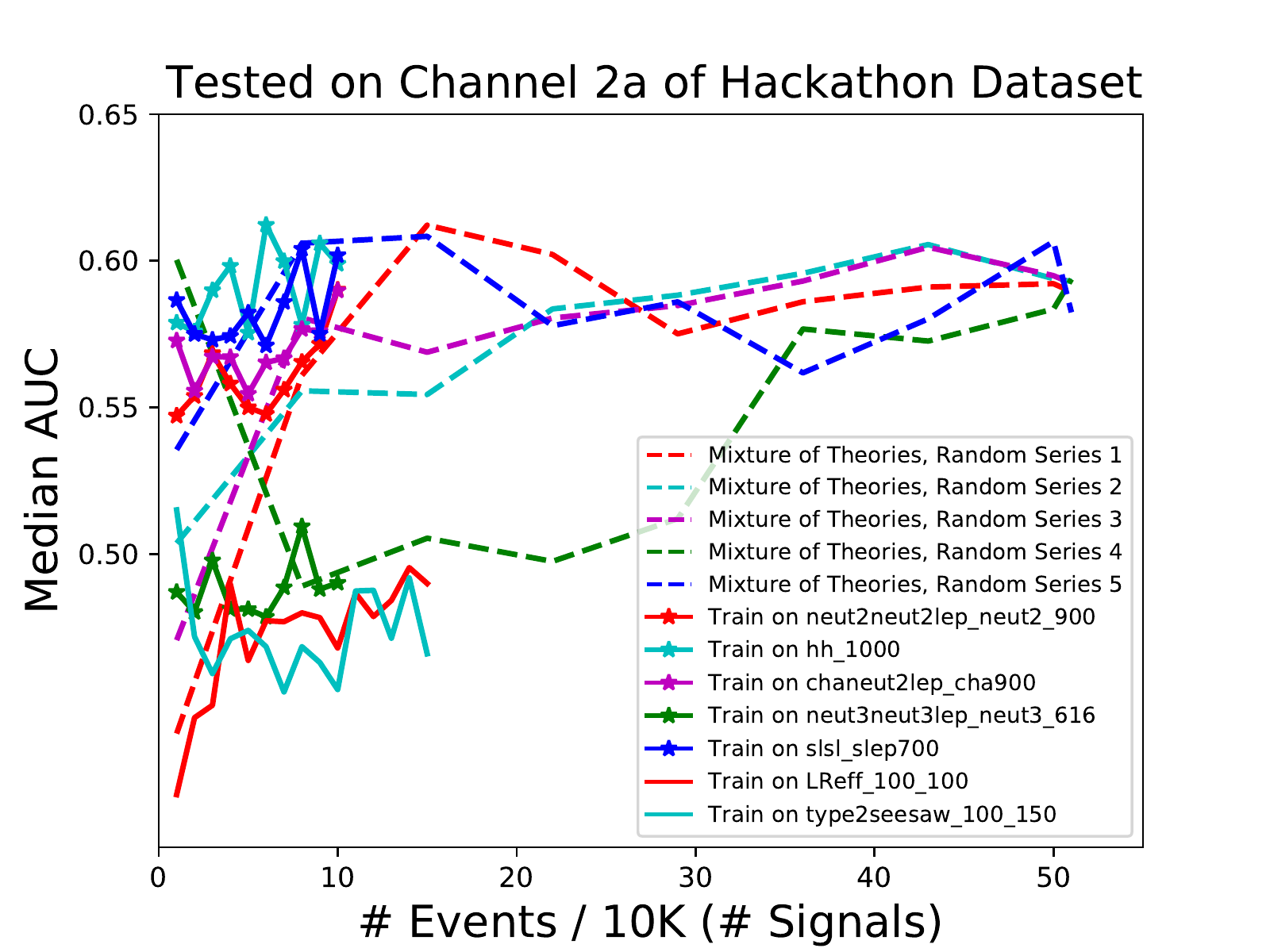}
    \includegraphics[width=0.8\textwidth]{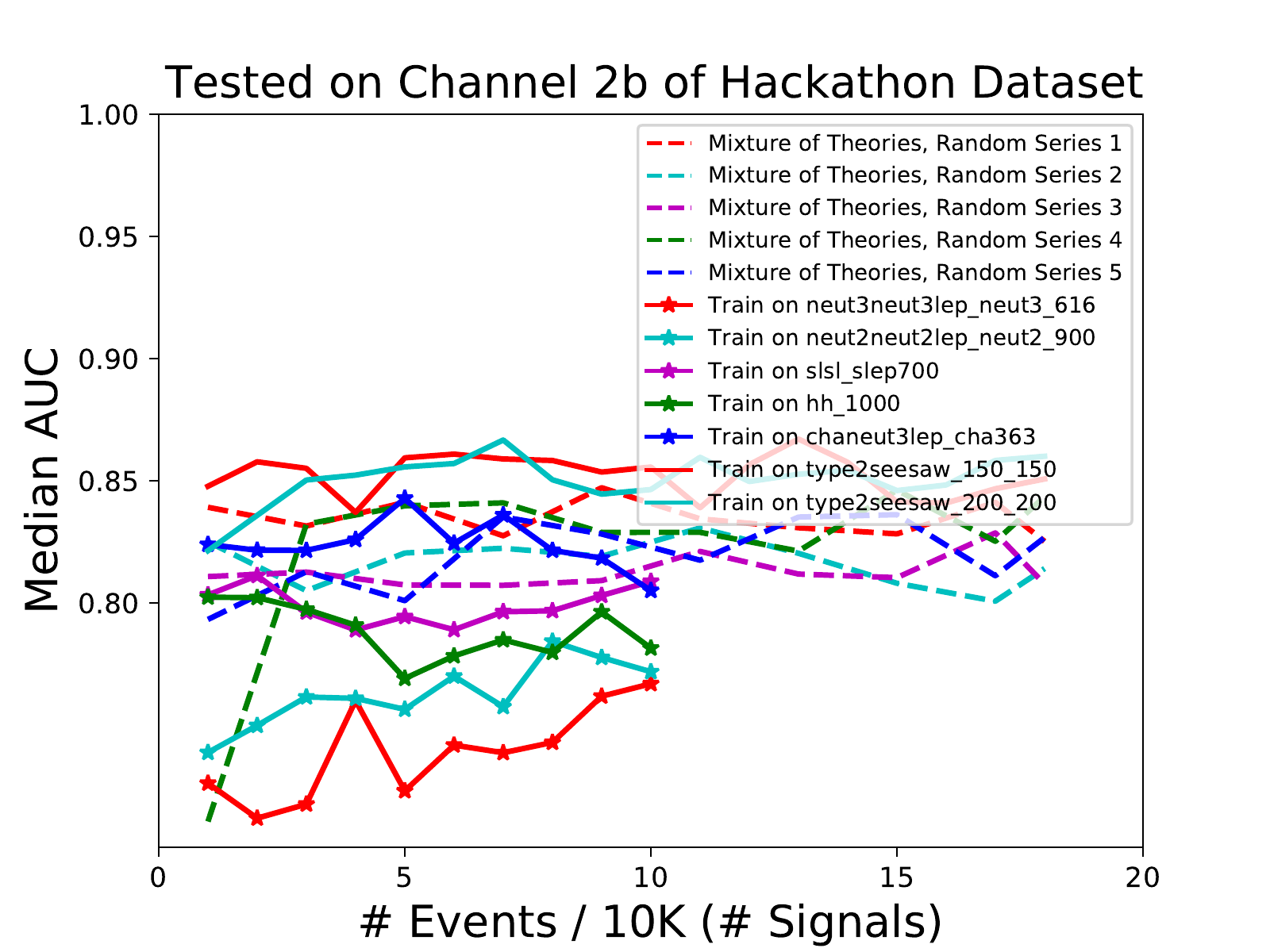}
    \caption{The figure shows the median AUCs of the trained ML models for channels with leptonic final states, tested on signals from the {\tt hackathon data}. For the dashed line, the training processes based on an increasing number of signals with a point size of 10k are repeated 5 times per channel in random series. The steps to add signals to the training set are (7, 2) for (channel 2a, channel 2b), while all channels start with the ML model with only one signal in the training set. For the star--shaped lines we use only 1 single signal and increase the number of events gradually in steps of 10k. The solid lines are similar to the star lines trained with a single signal. However, additional events are generated to extend the line.}
\label{fig:AUC2}
\end{figure}

\begin{figure}[!htbp]
    \center
    \includegraphics[width=0.8\textwidth]{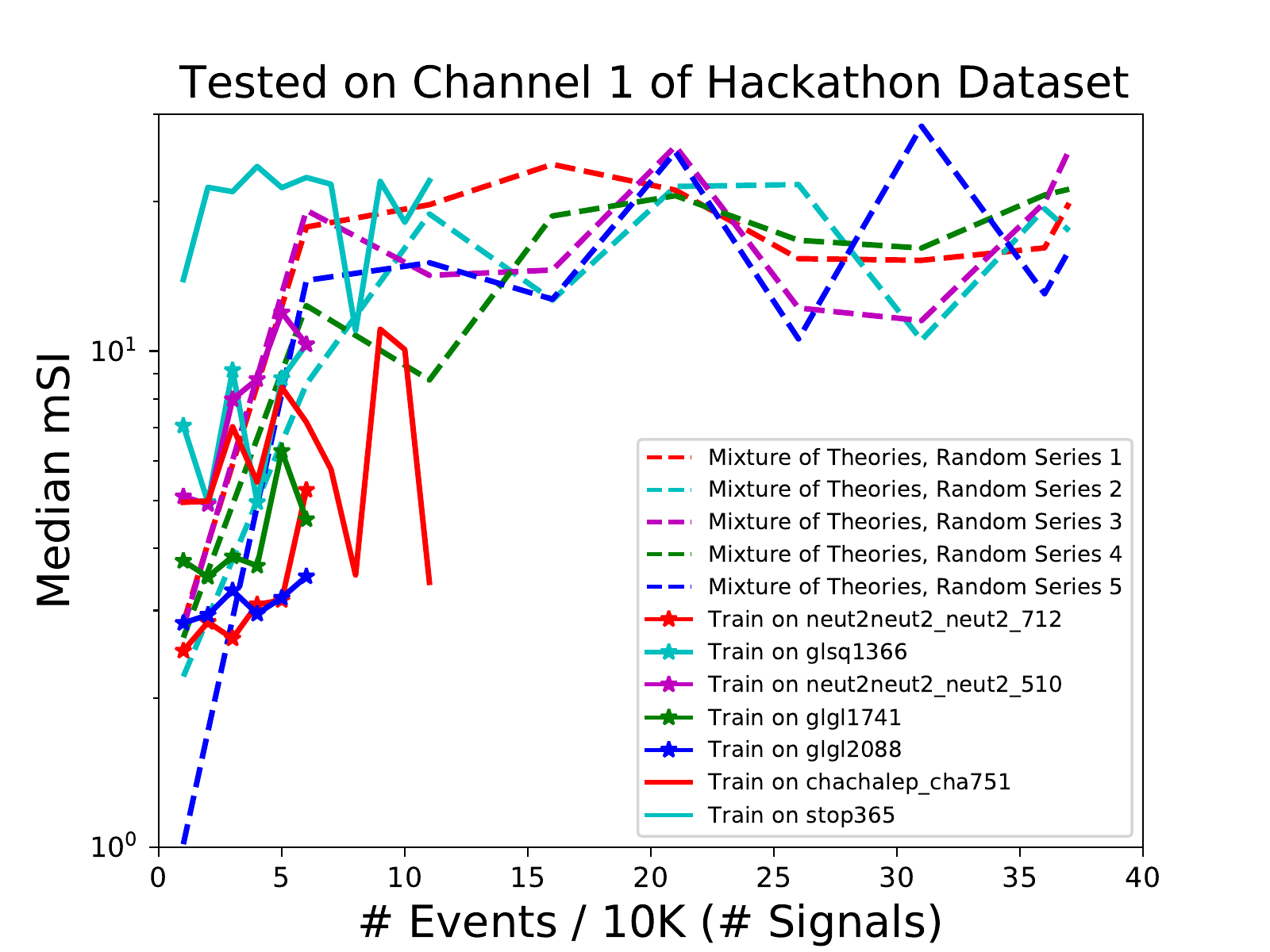}
    \includegraphics[width=0.8\textwidth]{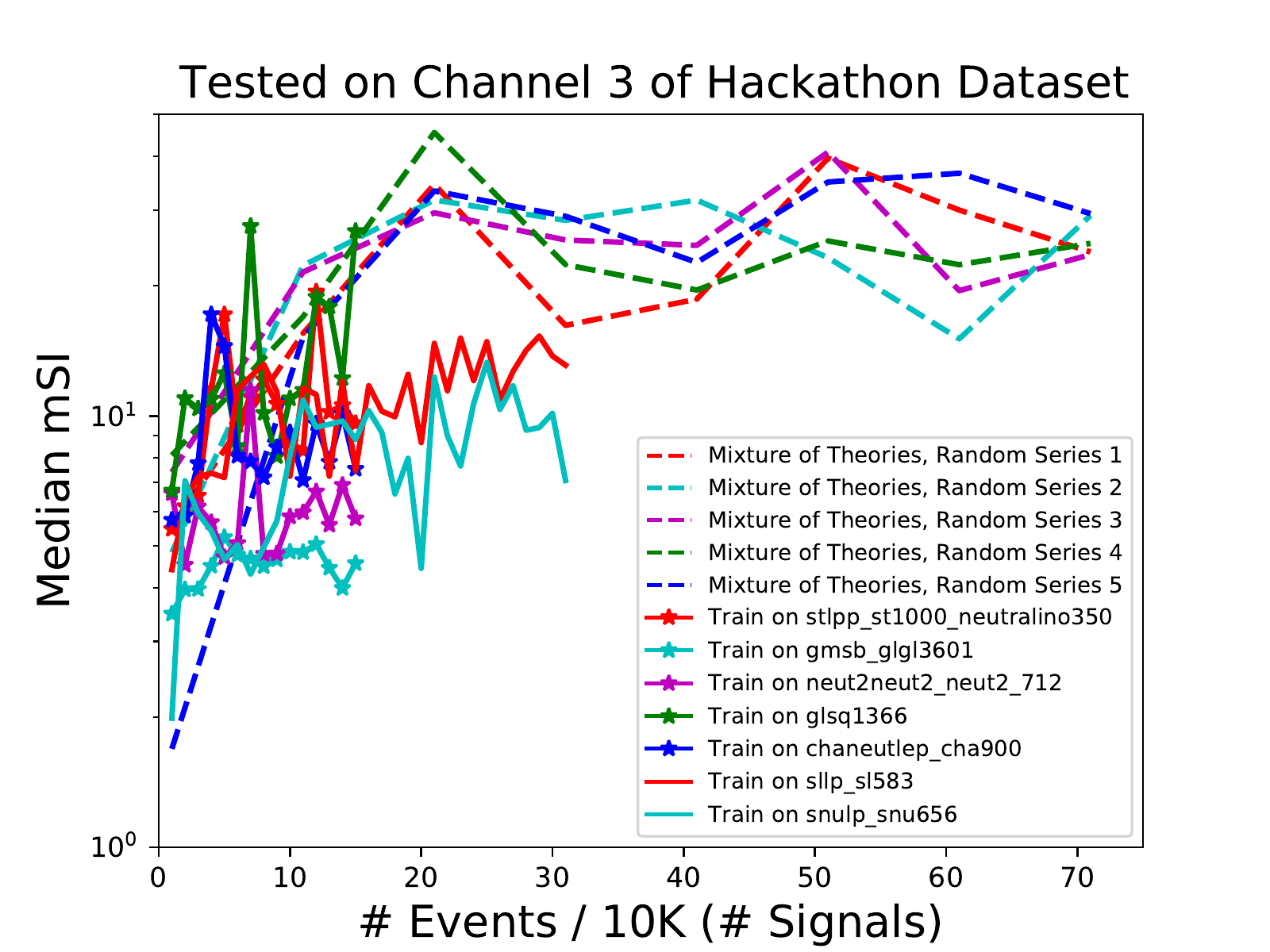}
    \caption{The Figure shows the median values of maximum significance improvements for all test signals from trained models for channels with hadronic final states, tested on signals from {\tt hackathon data}. For dashed line, the training processes based on increasing number of signals with point size equaling to 10000 are repeated 5 times per channel in random series. The steps of adding signals to training set are (5, 10) for (Channel 1, Channel 3) respectively, while all channels begin from the model with only one signal in the training set. For the star lines, we only use 1 single signal and gradually increase the event number with step equaling to 10000. The solid lines are similar with star lines trained on single signal. However, extra events are generated to extend the line.}
\label{fig:mSI1}
\end{figure}

\begin{figure}[!htbp]
    \center
    \includegraphics[width=0.8\textwidth]{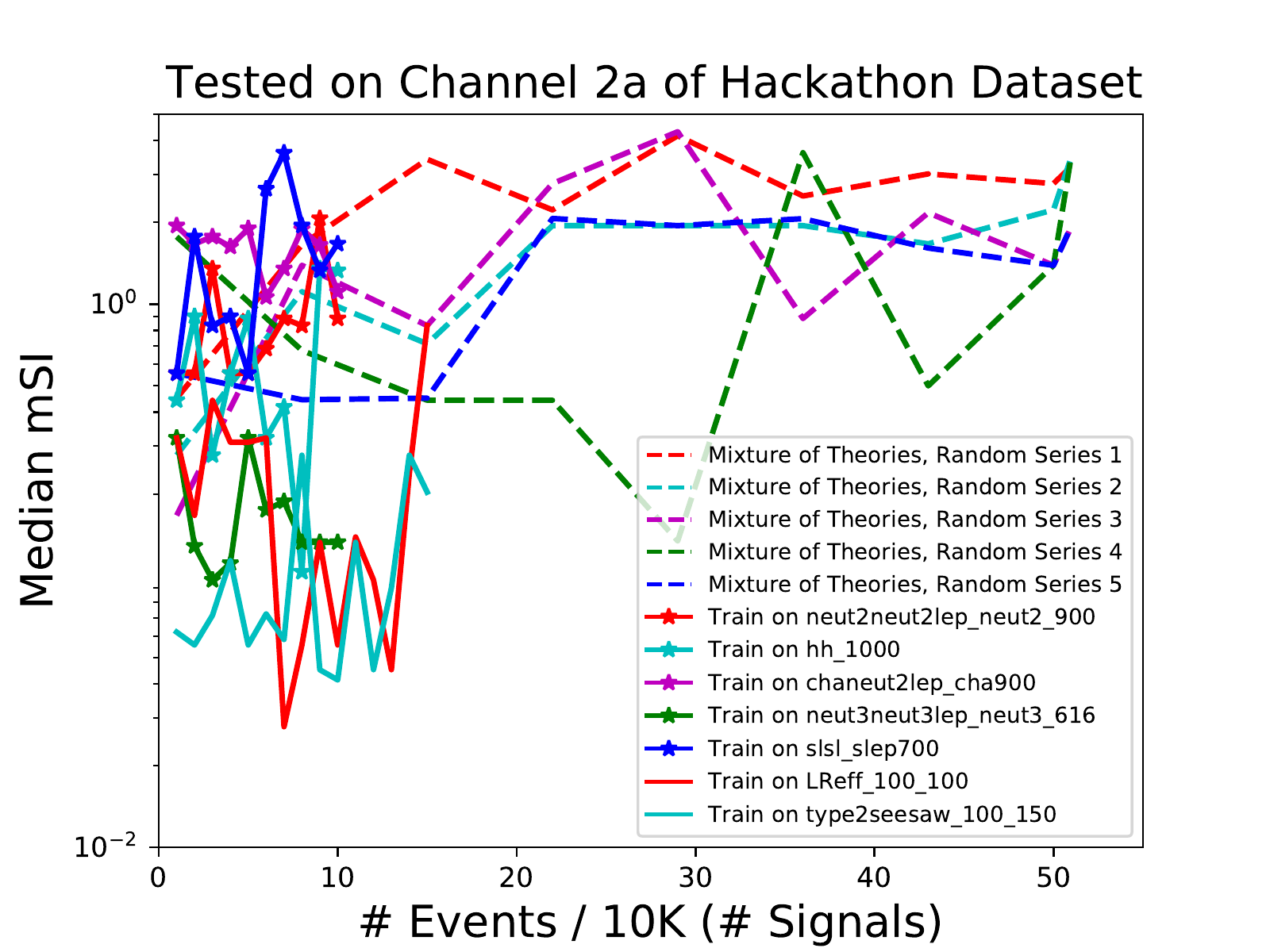}
    \includegraphics[width=0.8\textwidth]{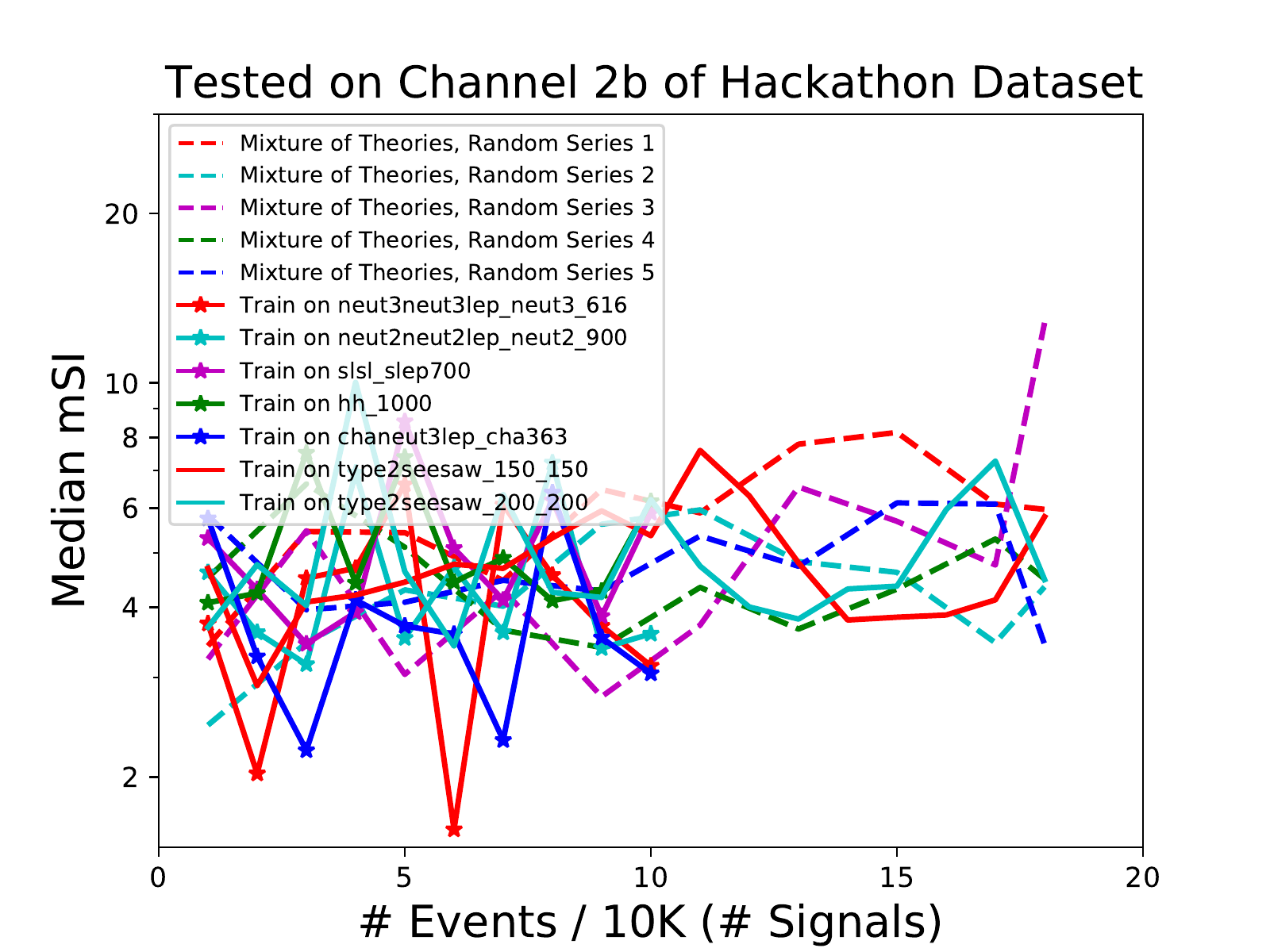}
    \caption{The figure shows the median values of the maximum significance improvements for all test signals from trained models for channels with leptonic final states tested with signals from the {\tt hackathon data}. For the dashed line, the training processes based on an increasing number of signals with a point size of 10000 are repeated 5 times per channel in random series. The steps to add signals to the training set are (7, 2) for (channel 2a, channel 2b), while all channels start with the model with only one signal in the training set. For the star--shaped lines, we use only 1 single signal and gradually increase the number of events in steps of 10000. The solid lines are similar to the star lines trained with a single signal. However, additional events are generated to extend the line.}
\label{fig:mSI2}
\end{figure}

\begin{figure}[!htbp]
    \center
    \includegraphics[width=\textwidth]{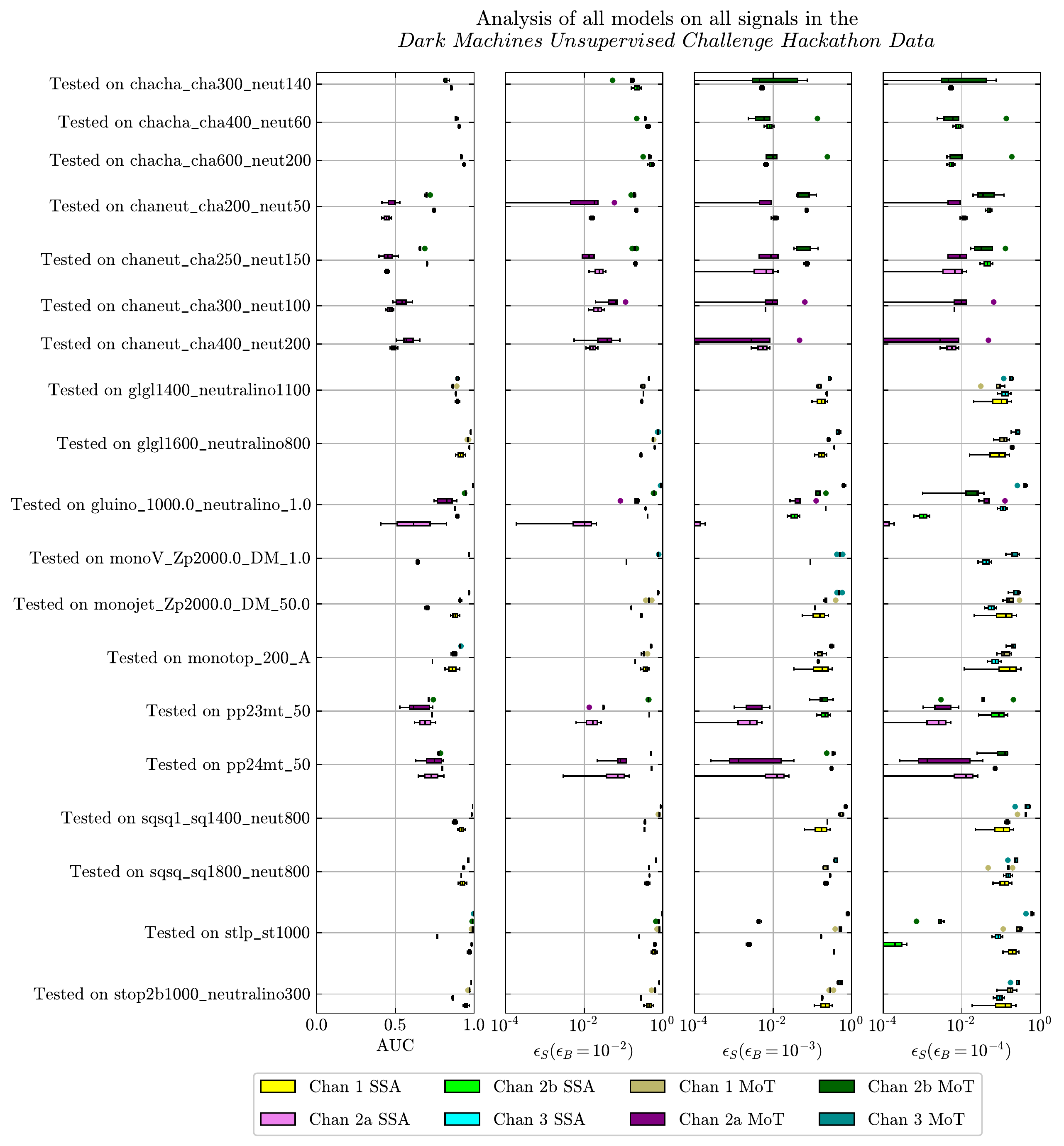}
    \caption{The figure shows the results for AUC and $\epsilon_S$ with several $\epsilon_B$ ($10^{-2}$, $10^{-3}$ and $10^{-4}$) tested with signals from the {\tt hackathon data}. This compares the supervised methods trained with a mixture of theories (boxes above the middle lines) to those trained with a single signal approach (boxes below the middle lines). Chan X SSA means the results for channel X from the single signal approach. Chan X MoT means the results for channel X from a mixture of theories. SSA contains two ML models corresponding to the endpoints of the solid lines in Fig.~\ref{fig:AUC1}, \ref{fig:AUC2}, \ref{fig:mSI1} and \ref{fig:mSI2}, while MoT contains 5 ML models corresponding to the points on the dashed lines that have the same event number as the endpoints of the solid lines.}
\label{fig:AUC3}
\end{figure}

\begin{figure}[!htbp]
    \center
    \includegraphics[width=\textwidth]{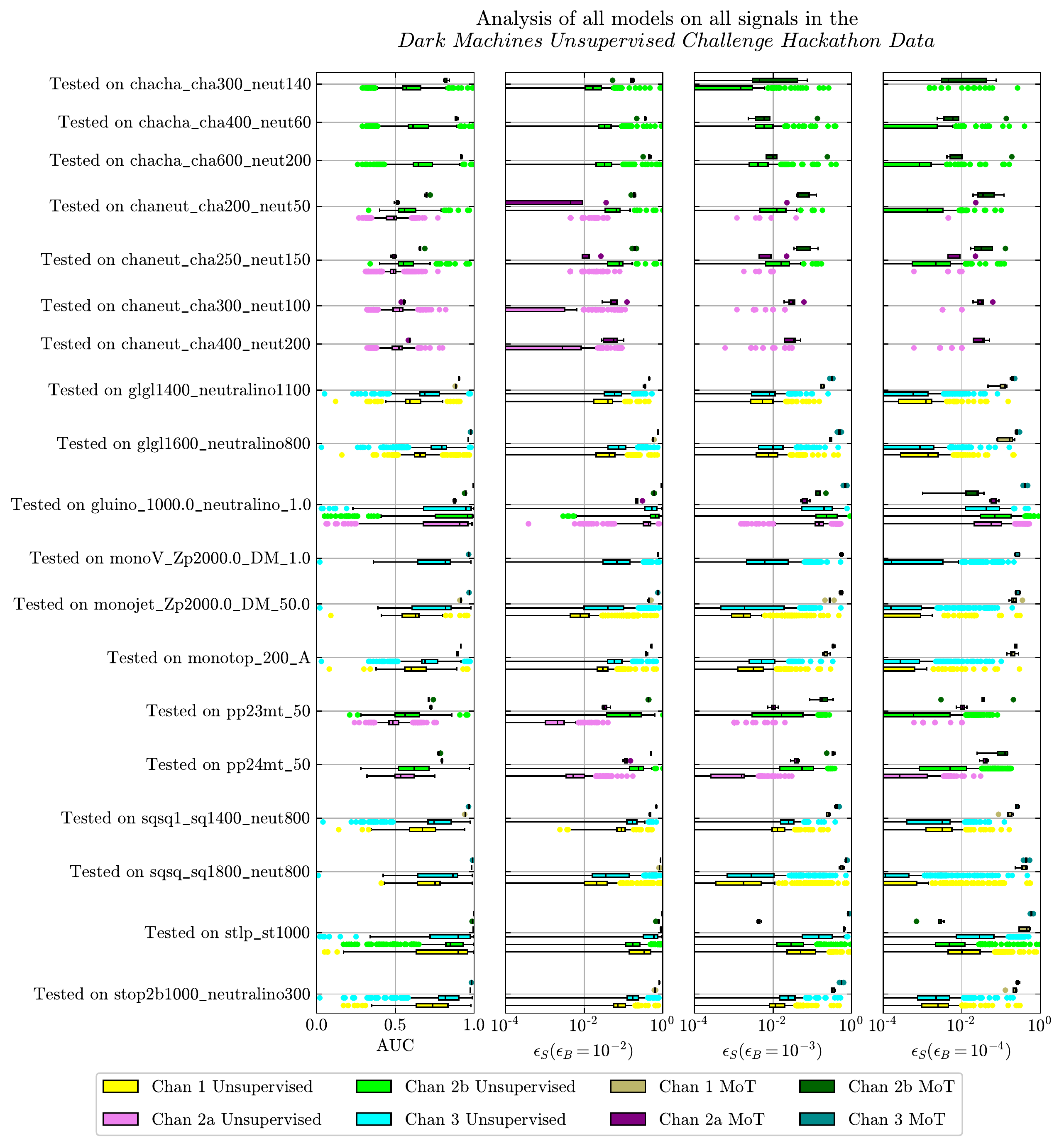}
    \caption{The figure shows the results for AUC and $\epsilon_S$ with multiple $\epsilon_B$ ($10^{-2}$, $10^{-3}$ and $10^{-4}$) comparing the supervised methods trained from the mixture of theories (boxes above the middle lines) and the unsupervised methods (boxes below the middle lines). Chan X Unsupervised denotes the results for channel X from unsupervised methods, while Chan X MoT denotes the results for channel X from supervised methods. For each individual channel, the ML models above the central line correspond to the 5 endpoints of the dashed lines in Figs.~\ref{fig:AUC1}, \ref{fig:AUC2}, \ref{fig:mSI1} and \ref{fig:mSI2}.}
\label{fig:AUC4}
\end{figure}

\begin{figure}[!htbp]
    \center
    \includegraphics[width=\textwidth]{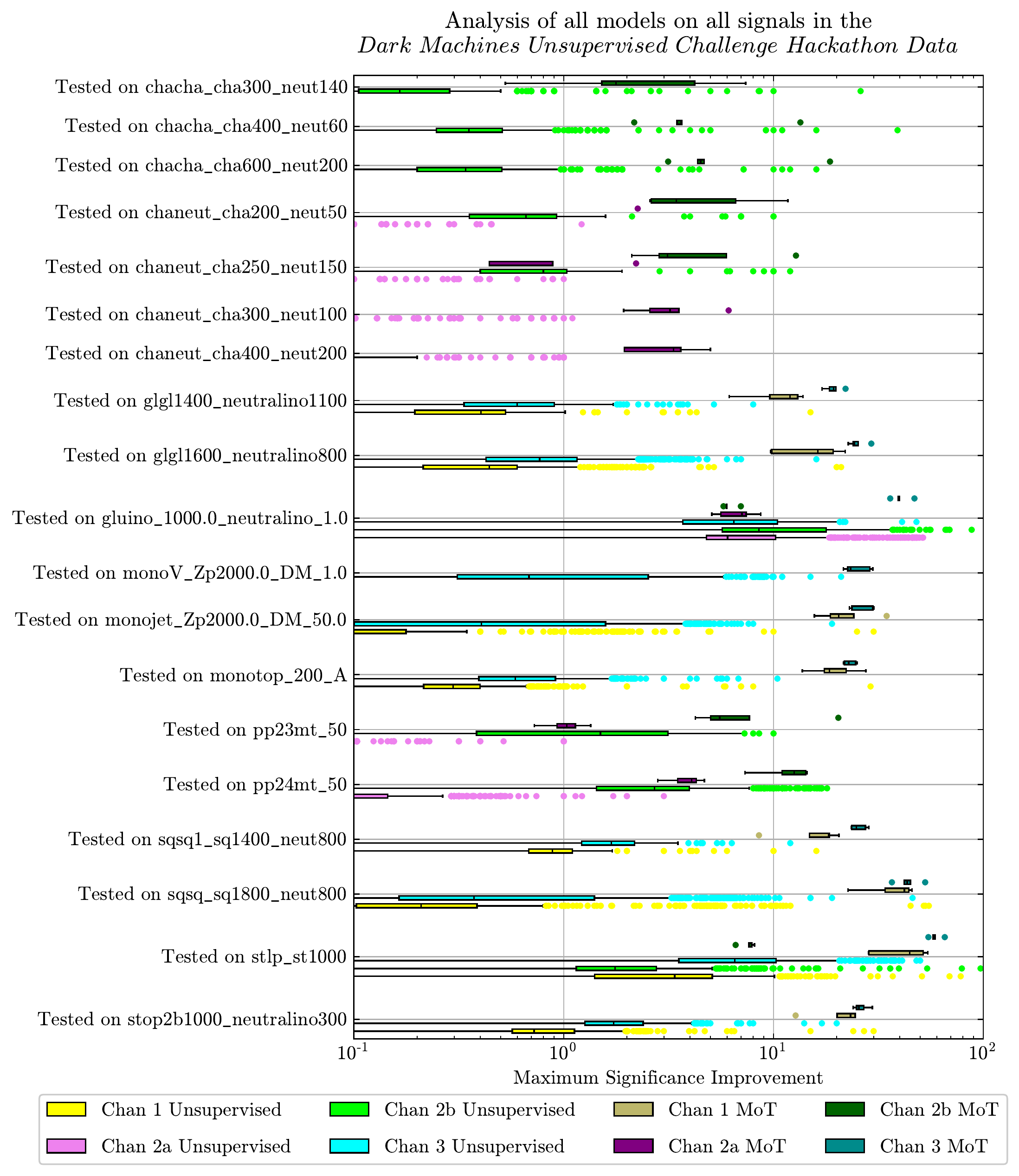}
    \caption{The figure shows the results for the maximum significance improvements comparing the supervised methods trained from a mixture of theories (above the central line) and the unsupervised methods (below the central line). Chan X Unsupervised denotes the results for channel X from unsupervised methods, while Chan X MoT denotes the results for channel X from supervised methods. For each individual channel, the ML models above the central line correspond to the 5 endpoints of the dashed lines in Figs.~\ref{fig:AUC1}, \ref{fig:AUC2}, \ref{fig:mSI1} and \ref{fig:mSI2}.}
\label{fig:mSI4}
\end{figure}

\FloatBarrier
\bibliographystyle{JHEP}

\end{document}